\documentclass[preprint]{aastex}

\usepackage[colorlinks,urlcolor=blue,citecolor=blue,linkcolor=blue]{hyperref}
\usepackage{graphicx}
\usepackage{natbib}
\usepackage{pdflscape} 
\usepackage{amssymb}
\usepackage{ulem} 
\usepackage{color} 

\newcommand{\be}{\begin{equation}}
\newcommand{\ee}{\end{equation}}
\newcommand{\nn}{\mbox{} \nonumber \\ \mbox{} }
\newcommand{\ba}{\begin{eqnarray}}
\newcommand{\ea}{\end{eqnarray}}
\newcommand{\om}{\omega}
\newcommand{\Alfven}{ Alfv\'{e}n }

\newcommand{\curl}{{\rm curl\, }}
\newcommand{\E}{{\bf E}}
\newcommand{\B}{{\bf B}}

\renewcommand{\v}{{\bf v}}

\newcommand\eg{{\it{{e.g.,\ }}}}

\newcommand{\Lf}{{Lorentz factor}}
\newcommand{\Bf}{{magnetic field}}
\newcommand{\Bfs}{{magnetic fields}}
\newcommand{\Ef}{{electric  field}}

\newcommand{\NS}{neutron star}

\newcommand{\BH}{{black hole}}

\newcommand{\Fermi}{{\it Fermi}}

\begin{document}

\title{Self-similar structure of ultra-relativistic  magnetized termination shocks and the role of reconnection in long-lasting GRB ouflows}

\author{Maxim Lyutikov\\
Department of Physics, Purdue University, \\
 525 Northwestern Avenue,
West Lafayette, IN
47907-2036 }

\begin{abstract}
We consider   the double-shock structure of ultra-relativistic flows produced by the interaction of magnetized wind with  magnetized  external medium. 
The contact discontinuity (CD) is a special point in the flow - density, kinetic pressure and \Bf\ experience a jump or are non-analytic on the CD.  To connect dynamically the outside region (the forward shock flow) with the inside 
   region (the reverse shock flow) requires resolving flow singularities at the contact discontinuity. 
 On the CD the pressure is communicated exclusively by the \Bf\ on both sides - the  CD become an \Alfven  tangential  discontinuity. Thus,
the dynamic amplification of the \Bf\  leads to a formation of a  narrow  magnetosheath.
We  discussed a possibility  that particles emitting early $X$-ray afterglows,  as well as Fermi GeV photons, are accelerated via magnetic reconnection processes in the post-reverse shock  region, and in the magnetosheath in particular.
 \end{abstract}

\maketitle
\section{Introduction}

Many astrophysical phenomena, like pulsar winds,  jets in Active Galactic Nuclei (AGNe) and Gamma Ray Bursts (GRBs) involve interaction of ultra-relativistic  magnetized outflow with external weakly magnetized medium. In case of  quasi-steady sources (pulsars and AGNe) the
wind's (or the  jet's)  dynamic pressure  acting over time  drives   the region of interaction with external medium to large distances, and correspondingly non-relativistic (or weakly relativistic) expansion velocities. Interaction with relativistic pulsar wind with the non-relativistic wind environment has been  considered by \cite{KennelCoroniti84} (K\&C below) and  \cite{1987ApJ...321..334E}. GRBs are different: for approximately a day after the explosion  the forward shock remains highly relativistic. There are also observational indictions that the central source continues activity for this time, and probably even longer, \eg\   \cite{2007ApJ...671.1903C}.  Models of the central source    -  be it the \BH\ or \NS\  \citep{2011MNRAS.413.2031M,2013ApJ...768...63L} - prefer magnetized  winds.

\begin{figure}[h!]
 \centering
 \includegraphics[width=.99\columnwidth]{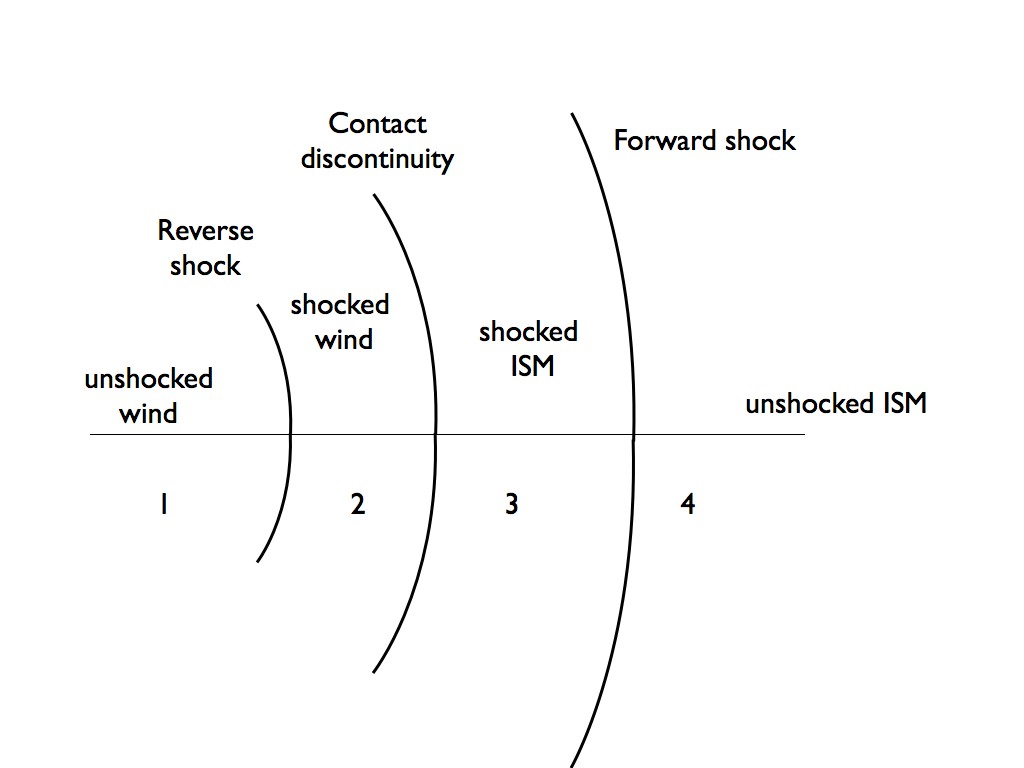}
 \caption{Cartoon of the model. The wind from the long-lasting central engine (region 1) passes through the reverse shock (RS) and forms a shocked wind flow (region 2) - the flow within this region is  the main goal of the present study. The wind drives a  forward  shock (region 3) into  ISM flow;  region 4 is the unshocked ISM.}
 \label{picture-RS}
\end{figure}

In this paper we consider the interaction of the relativistic magnetized wind with posibly magnetized external medium, Fig. \ref{picture-RS}. We assume that a  long-lasting central engine creates a  fast magnetized  wind that  drives a relativistic  shock into external medium (which can also be magnetized). At the same time, a reverse shock is launched in the wind. The main target of the present study is the structure of the shocked wind region  (region 2 in Fig. \ref{picture-RS}).

\section{Governing equations and boundary conditions}

Assuming a spherically symmetric outflow with toroidal magnetic field, the relativistic MHD equations 
  read \cite{Anile:1989}
\ba 
&&
\partial_t \left[ ( w+ b^2) \gamma^2 -(p+b^2/2)\right]+
{1\over r^2} \partial_r \left[ r^2 ( w+b^2) \beta \gamma^2 \right] =0
\label{x4} \\ &&
\partial_t \left[ ( w+ b^2) \gamma^2 \beta \right]+
{1\over r^2} \partial_r 
\left[r^2\left((w+b^2) \beta^2 \gamma^2 + (p+b^2/2)\right) \right] - 
{2 p \over r}=0
\label{x5} \\ &&
\partial_t \left[  b \gamma \right]+
{1\over r} \partial_r  \left[r b \beta \gamma \right] =0
\label{x51} \\ &&
\partial_t \left[ \rho \gamma \right]+
{1\over r^2 } \partial_r  \left[r^2 \rho  \beta \gamma \right] =0
\label{x6}
\ea
where  $w$ is the enthalpy, $b$ is the proper \Bf\ and other notations are standard. We choose to  work consistently with proper quantities, 
i.e. measured in the plasma
rest frame. One should be careful in comparing our equation with
 K\&C and  \cite{BlandfordMcKee} (B\&M below).

For highly relativistic outflows with \Lf\ $\gg 1$,
following   B\&M
we choose the self-similar variable
\be
\chi = [1+2(m+1) \Gamma^2](1-r/t)
\label{chi}
\ee
where   $\Gamma \propto t^{-m/2}$ is the Lorentz factor of the forward 
shock. For a wind power scaling as  $L_w \propto t^q$ and external density $ \propto r^{-k}$, $m=(2-q-k)/(2+q)$  (B\&M). Particularly interesting cases are constant energy source in constant density, $k=0,\, m=1$, and in external wind environment, $k=1, \, m=0$.

The following 
  parameterization (B\&M  and  \cite{2002PhFl...14..963L})
\ba &&
p = (2/3) \Gamma^2 n_1 m c^2 f(\chi)
\nn &&
\gamma^2 = \frac{1}{2} \Gamma ^2 g(\chi)
\nn &&
n= 2 \sqrt{2} \Gamma n_1 n(\chi)
\nn &&
b = \Gamma  b_1 h(\chi)
\label{ssss}
\ea
with the boundary conditions
$ f(1) =1$, $ g(1) =1$, $ n(1) =1$  eliminates $\Gamma(t)$ in the self-similar equations (in the fluid case, $b_1=0$; the case of  magnetized external medium was considered by  \cite{2002PhFl...14..963L}; this changes the normalization constants in Eq. (\ref{ssss})).

 Assume that the external plasma is weakly magnetized, so that the flow between the forward shock (FS) and the contact discontinuity (CD) is mostly  described by the  B\&M solution. (This will not be true near the CD even for infinitesimally small \Bf\ in the circumburst medium, see  \cite{2002PhFl...14..963L} and \S \ref{outs}.)
In the self-similar coordinate $\chi$  the MHD equations  (\ref{x6})
read
\cite{2002PhFl...14..963L}
\ba
&&
{{\cal{A}} \over g} { \partial \ln f \over  \partial \chi}= 
\frac{3 h^2 (3 g (m-2) \chi -4 m+4)-2 f (g \chi -2) (g (m-4) \chi -8 m+8)}{(m+1) (g
   \chi -2)}
   \nn &&
{{\cal{A}} \over g} { \partial \ln g \over  \partial \chi}= 
\frac{9 h^2 (m-1)-2 f (g (m+2) \chi -7 m+4)}{m+1}
\nn &&
{{\cal{A}}\over g} { \partial \ln h  \over  \partial \chi}= 
\frac{9 h^2 (m-1) (g \chi -2)-2 f
   \left(g^2 (m-4) \chi ^2+g (8-11 m) \chi +22 m-16\right)}{2 (m+1) (g \chi -2)}
\nn &&
{{\cal{A}}\over g} { \partial \ln n \over  \partial \chi}=
\frac{9 h^2 (g (m-3) \chi -2 m+2)-2 f \left(g^2 (m-6) \chi ^2+g (24-11 m) \chi +22 m-24\right)}{2
   (m+1) (g \chi -2)}
   \nn &&
{\cal{A}}  = 2 f \left(g^2 \chi ^2-8 g \chi +4\right)-9 g h^2 \chi
\label{15}
\ea

It is convenient to change to  a new variable $x= g \chi$,

\ba
&&
{{\cal{A}} } { \partial \ln f \over  \partial x}= 
\frac{2 f (x-2) (m (x-8)-4 x+8)+3 h^2 (m (4-3 x)+6 x-4)}{2 (x-2)}
   \nn &&
{{\cal{A}} } { \partial \ln g \over  \partial x}= 
\frac{1}{2} \left(2 f (m
   (x-7)+2 (x+2))-9 h^2 (m-1)\right)
\nn &&
{{\cal{A}}} { \partial \ln h  \over  \partial x}= 
-\frac{9 h^2 (m-1) (x-2)-2 f \left(m \left(x^2-11 x+22\right)-4
   \left(x^2-2 x+4\right)\right)}{4 (x-2)}
\nn &&
{{\cal{A}}} { \partial \ln n \over  \partial x}=
\frac{2 f \left(m \left(x^2-11 x+22\right)-6 (x-2)^2\right)-9 h^2 (m (x-2)-3 x+2)}{4
   (x-2)}
   \nn &&
  {{\cal{A}}}= 
f \left(m (x-4)+x^2+12 x-4\right)+9 {h}^2 x
\label{16}
\ea


Eqns (\ref{16}) describe  the double-shock structure of the magnetized wind-external medium environment  in self-similar coordinates chosen to match the unmagnetized part of the FS region. Our goal is to extend the solutions to the magnetized shocked-wind part of the flow. Clearly, the contact discontinuity, located at $x=2$ is a  special point of the equations (for density, kinetic pressure and \Bf, but not for the \Lf).  This creates, in the case of non-zero \Bf, a mathematically subtle problem. Resolving the behavior of the flow functions at this special point is the main topic of the paper.

\section{Magnetized wind  flow}

\subsection{Can the flow between RS and the CD be self-similar?}
\label{self}

The flow between the FS and CD is self-similar, as established by B\&M. In case of power supplied continuously  by the central source, should the corresponding flow between the RS and the CD be self-similar as well? This is a subtle question. First, 
the  solutions in the RS region cannot be resolved starting from the RS in a way similar to the FS case.
In the  case of the FS, the post-shock pressure, density and the \Lf\ scale as $p \propto n_{ext}  \Gamma_{FS}^2, n  \propto n_{ext}  \Gamma_{FS}, \gamma_{post-shock}  \propto \Gamma_{FS}$; expansion in $\Gamma_{FS} \gg 1$ then leads to the self-similar  B-McK solution. In the case of the reverse shock the scaling are strikingly different. If  reverse shock (RS) moving with $\Gamma_{RS} (t)$,   the wind's \Lf\ is  $\gamma_w$ (all in observer frames), wind rest-frame density is $n_w$ and the post-shock velocity in the frame of the RS is $\beta_{RS}'$ (see Eq. (\ref{gammaRS})),
 then 
the post-RS  scalings are
\ba &&
p_{post-RS} \propto  \left( \frac{\gamma_w}{\Gamma_{RS}} \right)^2 n_w
\nn &&
n_{post-RS}  \propto \frac{\gamma_w}{\Gamma_{RS}} n_w
\nn && 
\gamma_{post-RS} \propto \Gamma_{RS} \sqrt{ \frac{1+\beta_{RS}' }{1-\beta_{RS}'}} 
\ea
The above scalings, especially for $\gamma_{post-RS}$,  are drastically different from the FS case.
This implies that the self-similar solutions cannot be dexrived starting from the RS - {\it  it is the connection to the self-similar flow in the FS region that makes the flow in the RS self-similar.}
Thus,  in order to calculate the structure of the RS region we need to understand how to pass the special point  - the CD.

\subsection{Connecting FS and RS flow - passing through the special point at the contact discontinuity}
Let us connect the unmagnetized flow  on the  outside  of the CD  with the shocked magnetized wind inside of the CD.
In the non-relativistic case this was discussed by  \cite{1976PhFl...19.1889R} 
 (we denote ``outside'', with $_-$ subscript, the outside part corresponds to $x<2$), and ``inside'', with $_+$ subscript,  as a RS flow). 
In the fluid case the self-similar variable can be extended to the RS region, yet the  density profile  does not need to connect between the RS and FS regions. In the magnetized case then the question is how magnetic and kinetic pressure on the inside of the CD,  at $x=2_+$ combine to match the kinetic pressure on the outside, at $x=2_-$?

The two fluids on both sides of the CD should be in pressure balance and continuous velocity across the CD, but density, magnetization and kinetic pressure  can have a jump. 
On the inside (in the RS region), the equations (\ref{16})  close to the CD simplify
\ba
&&
 { \partial \ln f \over  \partial x}= 
 \frac{3 h^2 (m-4)}{2 (x-2) \left(f
   (m-12)-9 h^2\right)} 
     \nn && 
 { \partial \ln g \over  \partial x}= 
\frac{2 f (5 m-8)+9 h^2 (m-1)}{4 \left(f (m-12)-9 h^2\right)} 
 \nn &&
{ \partial \ln h  \over  \partial x}= 
\frac{
   (m-4) f }{(x-2) \left(9 h^2-f (m-12)\right)}
\nn &&
 { \partial \ln n \over  \partial x}=
\frac{2 f m+9 h^2}{(x-2) \left(18 h^2-2 f (m-12)\right)}
\label{18}
\ea
The pressure balance and continuity  require
\ba &&
f_-= f_+ + h^2 _+/2
\nn &&
g_-= g_+
\label{f4}
\ea
where $f_-$ and $g_-$ are given by the fluid solutions (\ref{solutions}) evaluated at $x=2$.

The system (\ref{18})  has an  integral of motion (if $k+m \neq 4$)
\be
f_+ +{3 h_+^2/4}= (3/2) f_{-}(2)
\label{fff1}
\ee
where $ f_{-}$ is the kinetic pressure in the unmagnetized outer part.

If we assume that on the inside boundary a fraction $\alpha$ of pressure is contributed by kinetic pressure, 
\ba &&
f_+= \alpha f_ -
\nn &&
h_+ = \sqrt{2 f_ - (1-\alpha)},
\ea
then condition (\ref{fff1}) and the pressure balance (\ref{f4}) require $\alpha=0$   - kinetic pressure must vanish on the inside of the CD.
 Thus, we demonstrated that  {\it  even a small  \Bf\  in the wind   fully balances the outside  pressure on the CD}. Similar result has been obtained in case of magnetized outside medium,  \cite{2002PhFl...14..963L}. Non-relativistic case has been discussed by \cite{1976PhFl...19.1889R,1987ApJ...321..334E}.

In order to continue the flow through the CD we need to find the behavior of the  flow functions near the CD, and then make a small step away (into the  inside part) from the CD.
By equating the corresponding powers for $x\rightarrow 2_+$  in (\ref{18}) we find
\ba &&
f_+ =f_0 (x-2)^{(4-m-k)/6}
\nn &&
n_+ =n_0 (x-2)^{(2-k)/4}
\nn &&
h\propto \sqrt{2 f_{0_-} -(4/3) f_+}
\label{111}
\ea
Thus,   for $m+k < 4$ the pressure $f$ is  zero inside of  the CD: all the pressure is magnetic. 
For the special case $k=m=2$,  we find $ f \propto h^{4/3}$, so that both kinetic and magnetic pressures on the CD must be non-zero, with the separation  between magnetic and kinetic 
pressures  depending on $\sigma_w$, Appendix  \ref{m2k2}. For the wind environment $k=2$ it is then required that $m< 2$.

Relations (\ref{111}) allow us to step away from the CD. The two new constants $f_0$ and $n_0$ are determined (at this point implicitly) by the properties of the RS and, in turn by the wind particle and magnetic fluxes. 

Let us summarize the conditions on the CD.  In magnetized case the \Lf\ remains continuos through CD, kinetic pressure on the inside (magnetized part) is zero, but its derivative diverges, \Bf\ on the inside balances the kinetic pressure on the outside. Thus, in magnetized case, two functions - $f$ and $n$ cannot be continued  through the CD; the \Bf\ is finite on the CD and can be integrated inwards.

 \subsection{Finding wind properties and location  of the RS in terms of integrations constants $f_0$ and $n_0$. }
 
Above, we have discussed that inside on the CD plasma pressure and density experience non-analytic behavior, so that they cannot be continued from the FS region. Relations (\ref{111}) allow us to step away from the CD. The two new constants $f_0$ and $n_0$ are determined (at this point implicitly) by the properties of the RS and, in turn by the wind particle and magnetic fluxes. 



 
 In the RS frame the    \Lf\ of the post-RS  flow is (K\&C)
 \be
 \gamma_{RS}' = \sqrt{1+\frac{8 \sigma_w ^2+10 \sigma_w +1}{16 (\sigma_w +1)}+\frac{\sqrt{64 \sigma_w ^2 (\sigma_w +1)^2+20 \sigma_w 
   (\sigma_w +1)+1}}{16 (\sigma_w +1)}} \approx
   \left\{ \begin{array}{cc}
   \frac{3}{2 \sqrt{2}} , & \sigma_w \ll 1\\
    \sqrt{ \sigma_w} , &  \sigma_w \gg 1
    \end {array}
    \right.
    \label{gammaRS}
   \ee
   where $\sigma_w$ is the wind magnetization; prime in  $ \gamma_{RS}'$ indicates that this is the \Lf\ of the post-RS flow in the frame of the RS.
   
   Using relations 
   \be
   \Gamma_{RS}^2 = \Gamma_{FS}^2/\chi_{RS} 
   \ee
  we find that  in the observer frame the post-RS flow satisfies
  \be
  \gamma_{post-RS} = \Gamma_{FS}  \frac{(1+\beta_{RS}') \gamma_{RS}'}{\sqrt{\chi_{RS}}} \equiv \Gamma_{FS}  \sqrt{ g(\chi_{RS}) /2}
  \ee
   Thus, RS is located at
   \be
x_{RS}=   g(\chi_{RS}) \chi_{RS} = 
2 \frac{1+\beta'_{RS}}{1-\beta'_{RS}}
  =
   \left\{
   \begin{array}{cc}
   4 ,& \sigma_w =0 \\
    8 \sigma_w,  & \sigma_w \gg 1
    \end{array}
    \right.
    \label{xxRS}
    \ee

\begin{figure}[h!]
 \centering
 \includegraphics[width=.99\columnwidth]{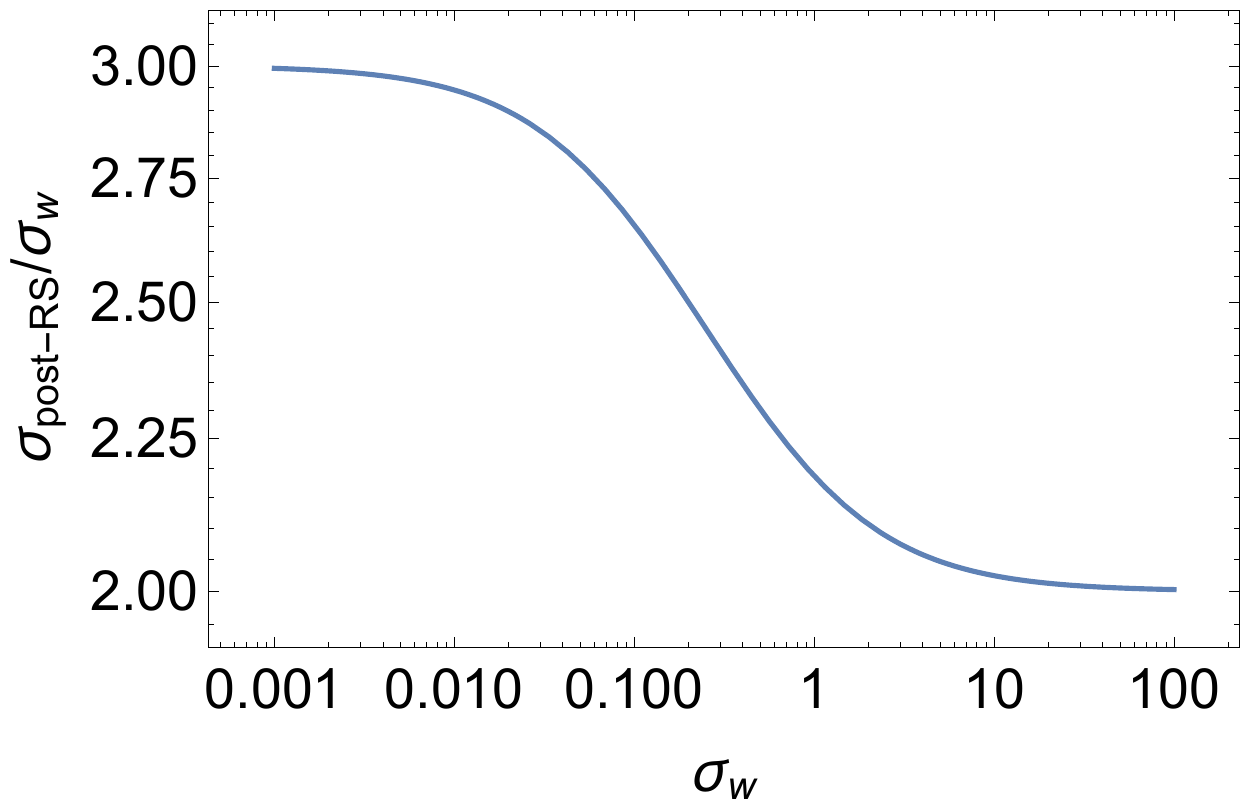}
 \caption{Post-RS magnetization as function of wind magnetization. 
}
 \label{sigmaRS1}
\end{figure}

Next, let the wind move in the laboratory frame  with \Lf\ $\gamma_w$ (and four-velocity $u_{w}$), have rest-frame \Bf\ $b_w$. 
Using magnetized shock conditions (K\&C), Eq (\ref{gammaRS}), the post-RS quantities are (denoted by subscript ${post-RS}$)
   \ba &&
   b_{post-RS} u_{post-RS} = b_w u_w
   \nn && 
   p_{post-RS} = \frac{ n_w  u_w^2}{
  4 \gamma_{RS}'   u_{RS}'}  \eta_\sigma
  \nn && 
\eta_\sigma  = \left(1 + \sigma_w (1 -   / \beta_{RS}')\right) 
  \nn &&
  \sigma_{post-RS} = \frac{ b_{post-RS}^2}{w_ {post-RS}}= \frac{\sigma_w}{\eta_\sigma}=
\left\{
\begin{array}{cc}
2 \sigma  & \sigma \gg 1
\\
3  \sigma  & \sigma \ll 1
\end{array},
\right.
\label{sigmaRS}
\ea
  see Fig. \ref{sigmaRS1}. Thus, for a given $\sigma_w$ we know the expected location of the RS (\ref{xxRS}) and the post-RS magnetization (\ref{sigmaRS}).

We then construct a numerical scheme that:
\begin{itemize}
\item for a given $m$ and $k$ calculates the thermal pressure and the \Lf\ in the unmagnetized portion of the shock, at $x=2_-$.
\item On the inside part of the shock uses scalings  (\ref{111}) with some $n_0$ and $f_0$ to step away towards inside, $x > 2$ from the  CD. 
This involves two constants $n_0$ and $f_0$; $n_0$ is just an overall normalization of density, but $f_0$ depends on the wind magnetization $\sigma_w$.
\item For a given  $f_0$ in the scaling of the kinetic pressure on the inside of the CD and given $h(CD_+)=\sqrt{ 2 f(CD_ -)}$ (balancing magnetic pressure n the inside and kinetic pressure on the outside the code integrates Eqns. (\ref{16}).
\item The local $\sigma$-parameter in the flow is give by $\sigma_{loc}= 3 h^2/( 8f)$.
\item We then calculate, inverting (\ref{sigmaRS})  what $\sigma_w$ would that local $\sigma_{loc}$ correspond to
\item Using the calculated $\sigma_w$ we estimate, using (\ref{xxRS}),  what at what   $x_{RS}$ that $\sigma_w$ should  be
\item The  procedure is run until the local $x_{RS}$ coincides with the expected  $x_{RS}$ calculate above.
\end {itemize}
When a convergence in the above procedure is reached - we have determined the value of the magnetization in the wind $\sigma_w$ and the location of the RS for an initial trail value of $f_0$.

\begin{figure}[h!]
 \centering
 \includegraphics[width=.32\columnwidth]{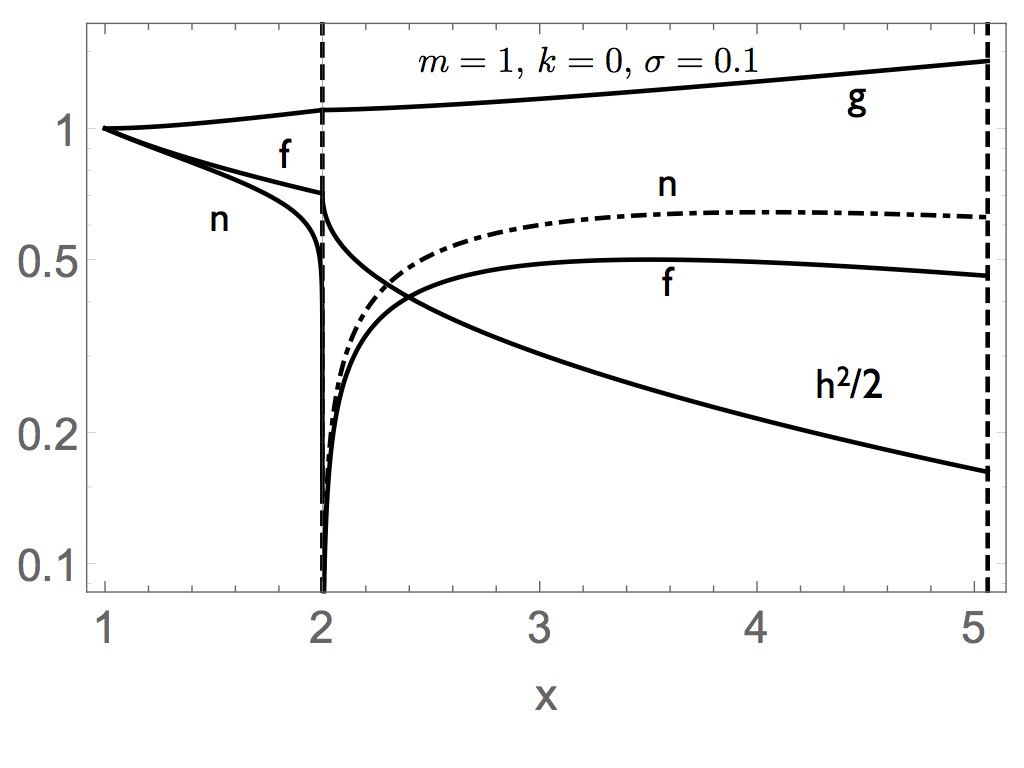}
  \includegraphics[width=.32\columnwidth]{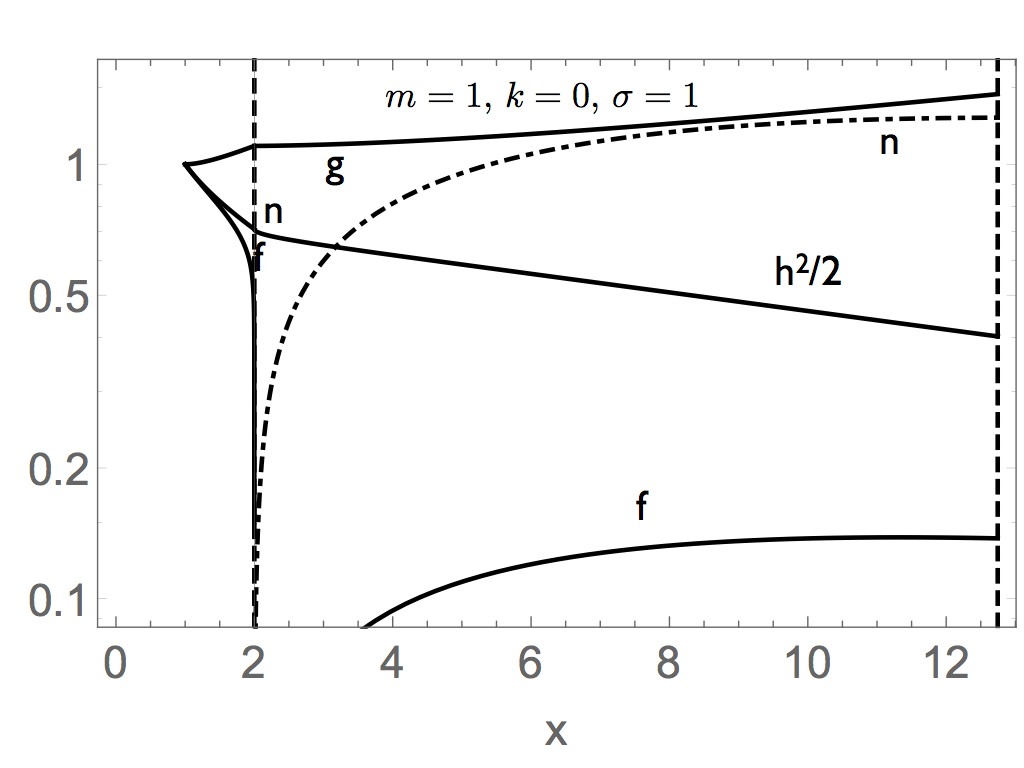}
   \includegraphics[width=.32\columnwidth]{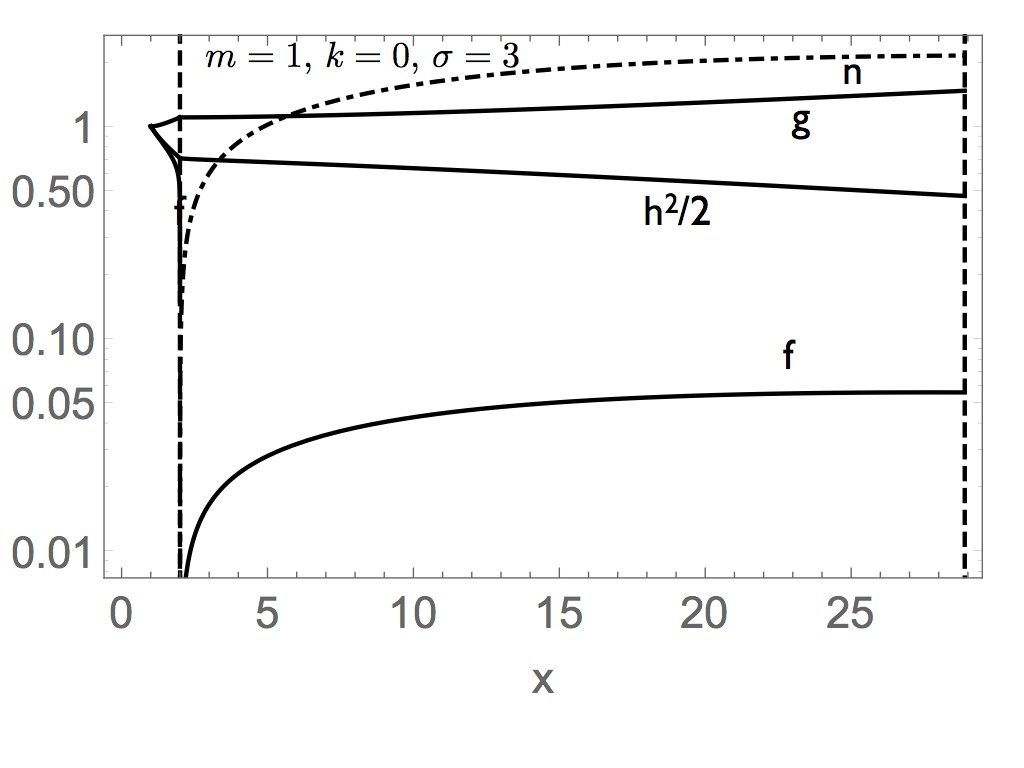}
   \\
    \includegraphics[width=.32\columnwidth]{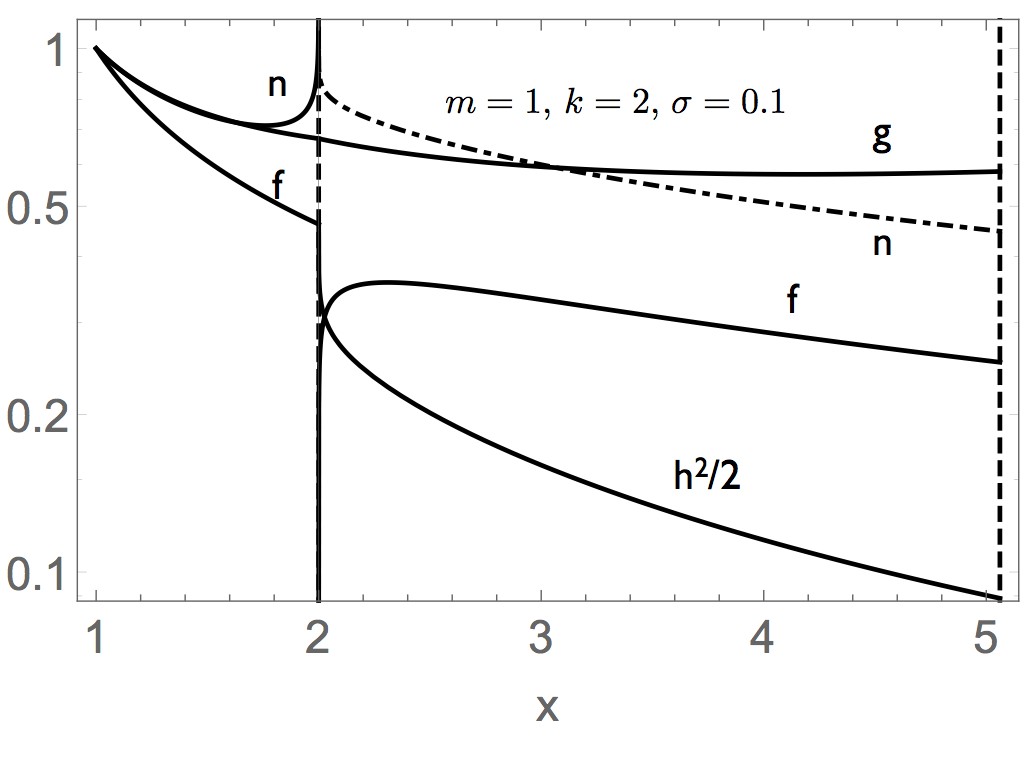}
     \includegraphics[width=.32\columnwidth]{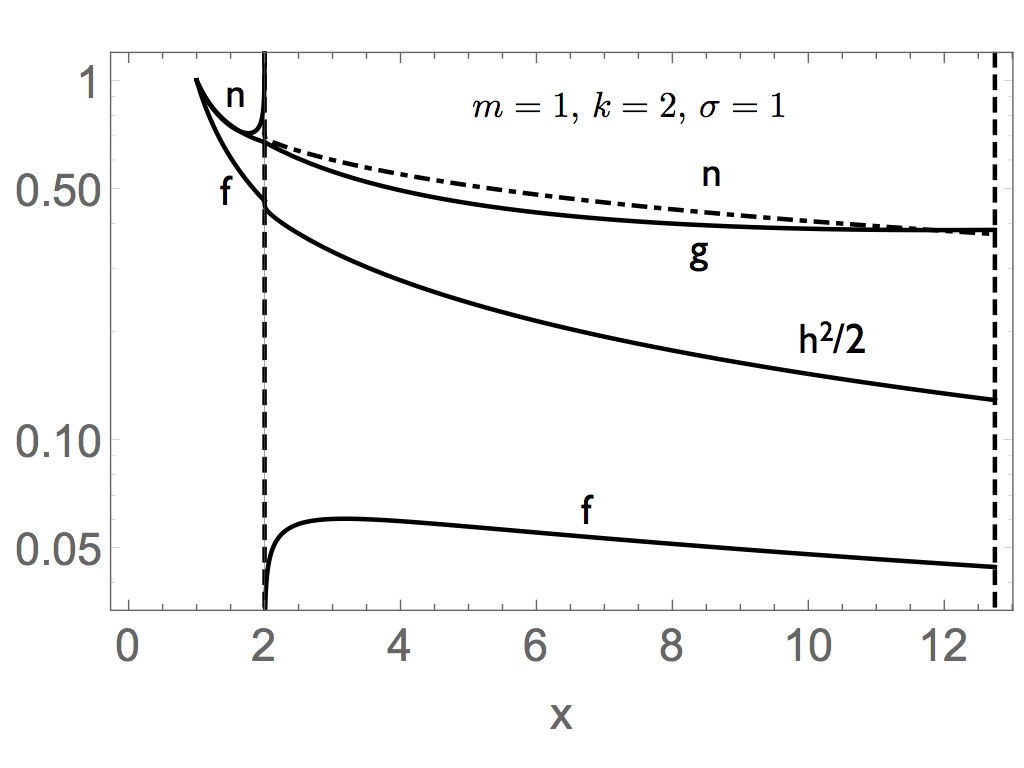}
      \includegraphics[width=.32\columnwidth]{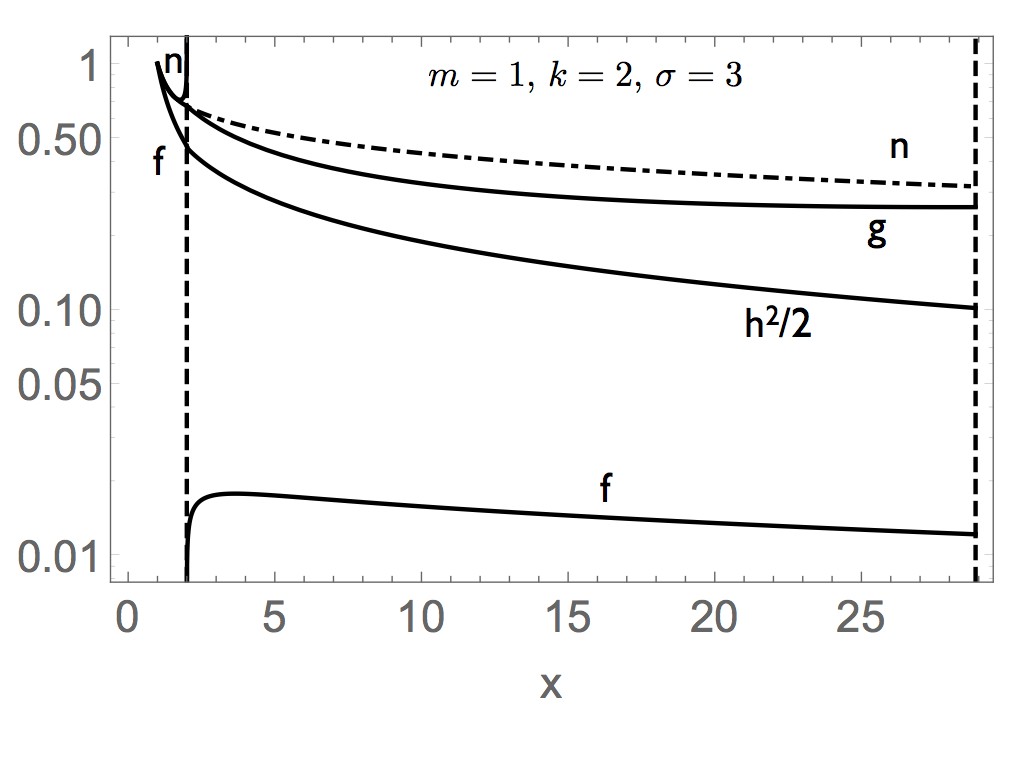}
      \\
       \includegraphics[width=.32\columnwidth]{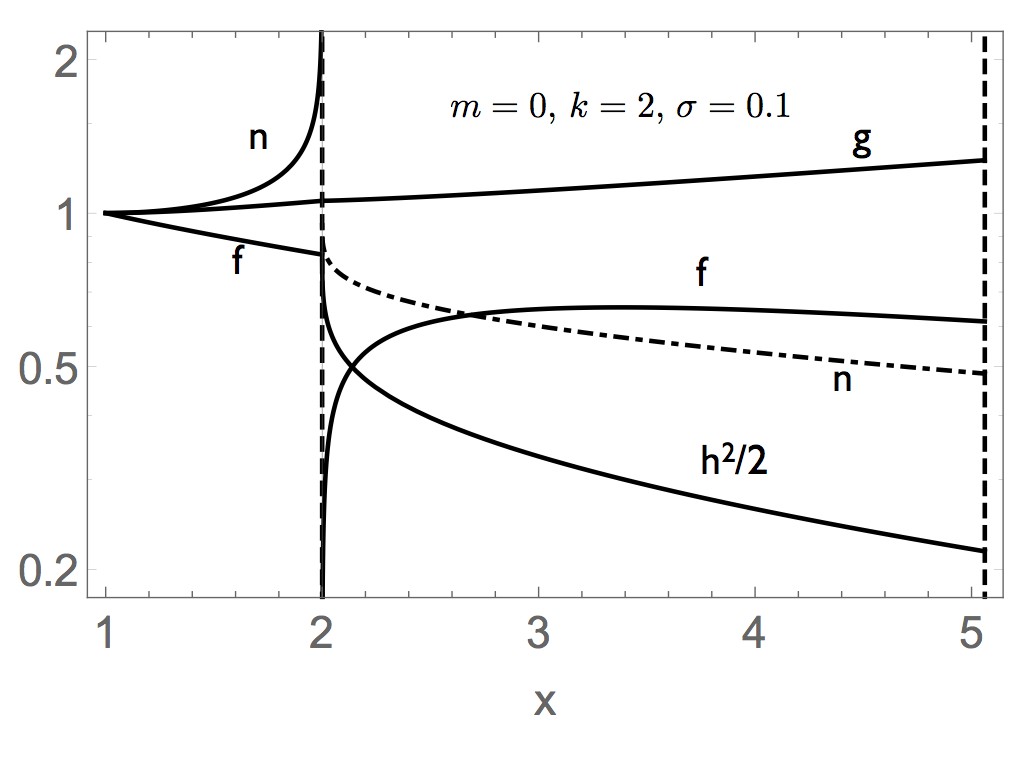}
        \includegraphics[width=.32\columnwidth]{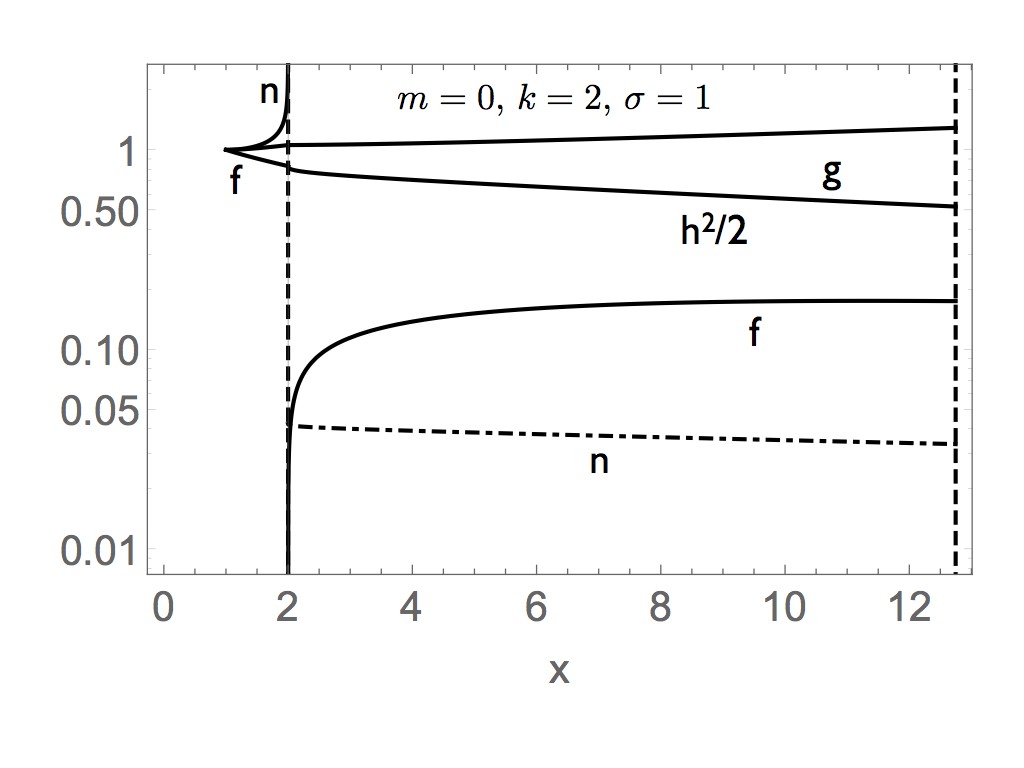}
         \includegraphics[width=.32\columnwidth]{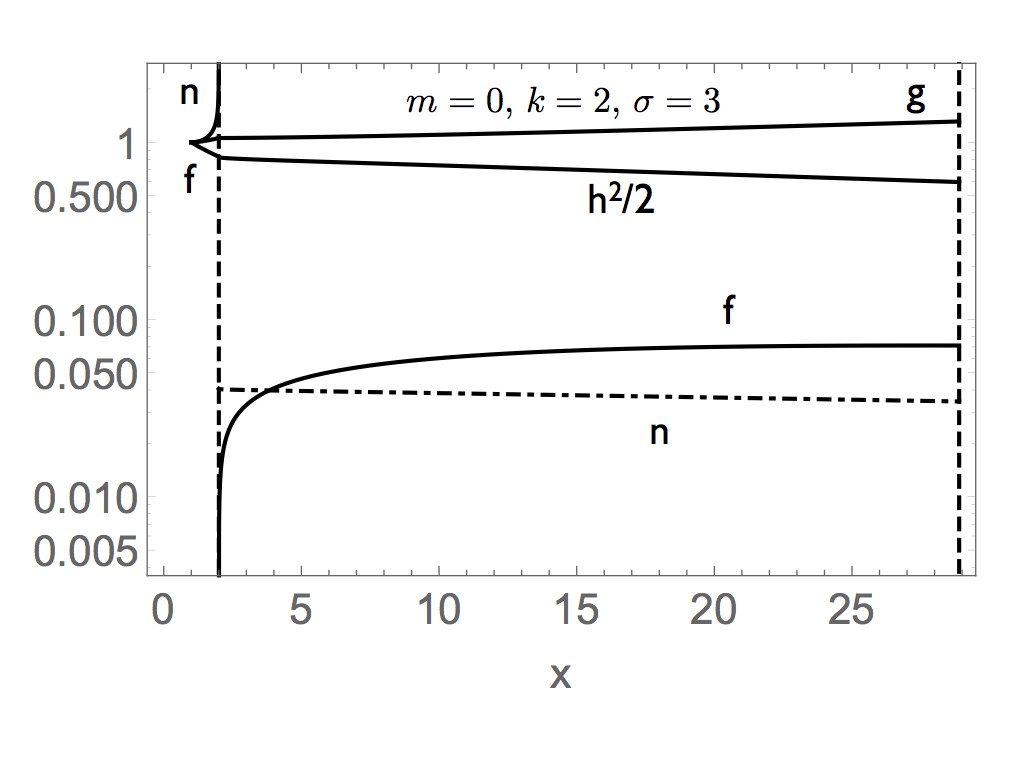}
          \\
       \includegraphics[width=.32\columnwidth]{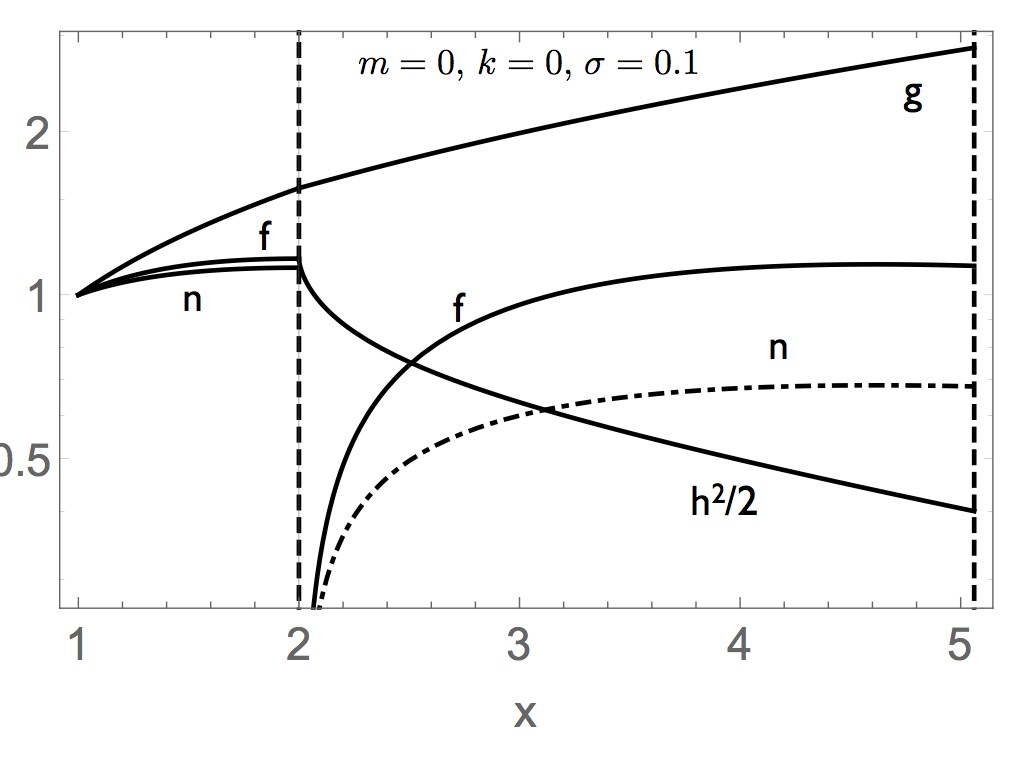}
        \includegraphics[width=.32\columnwidth]{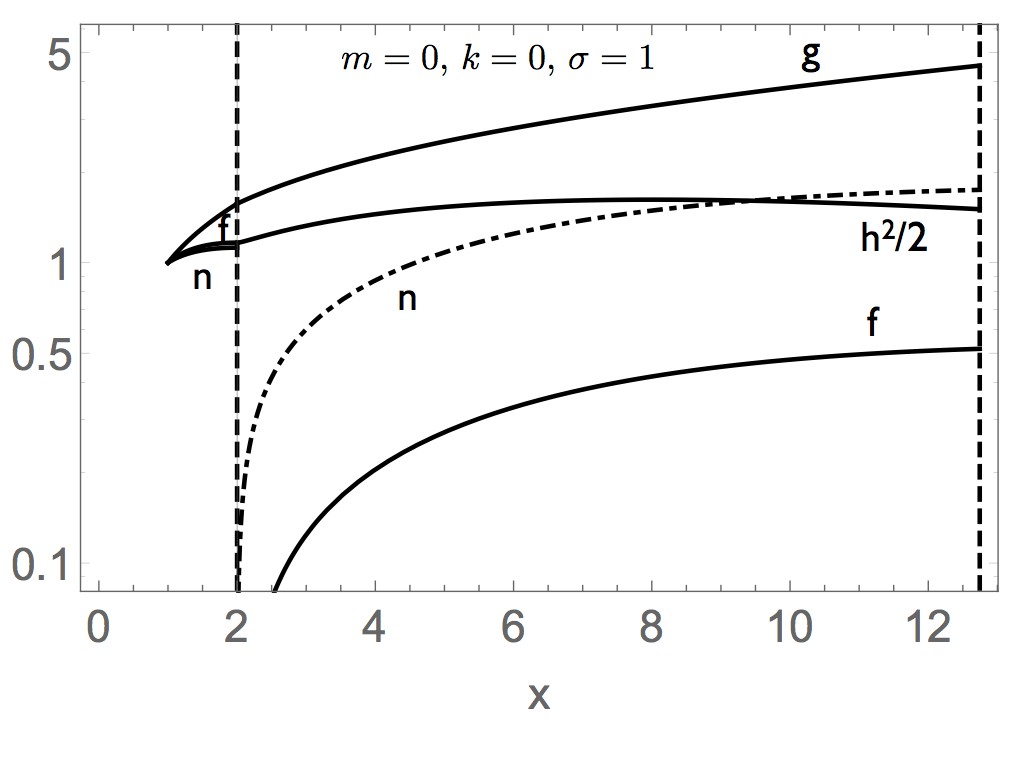}
         \includegraphics[width=.32\columnwidth]{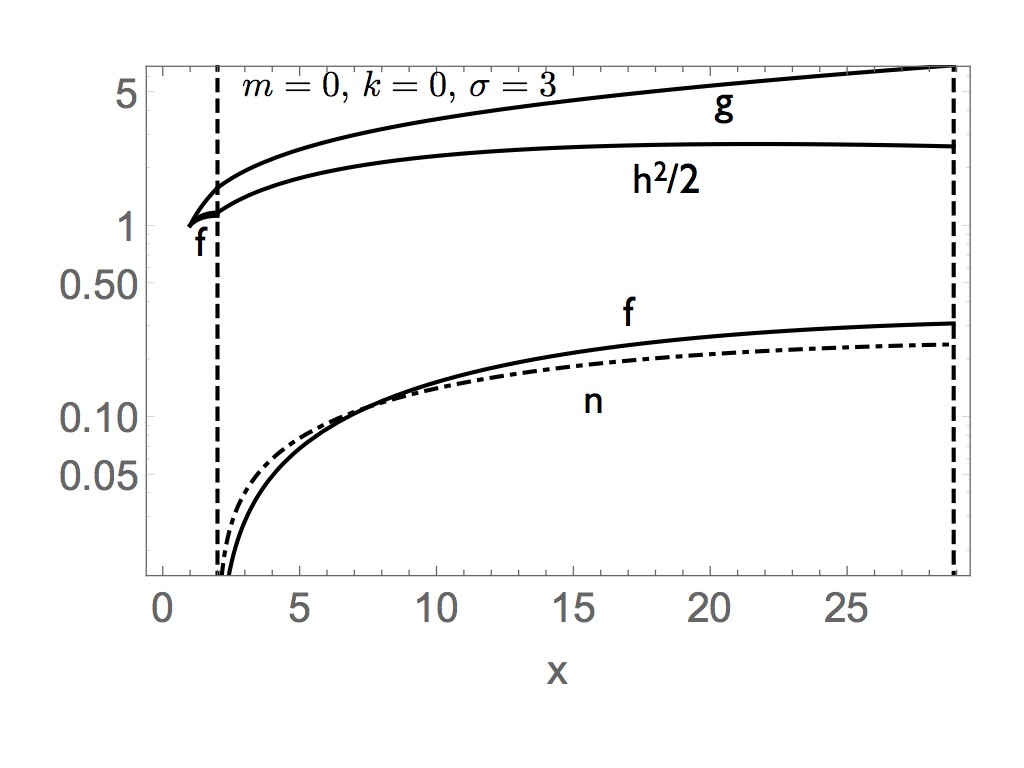}
 \caption{Self-similar structure of magnetized termination shock for different temporal profiles $m$, external density profiles $k$ and different wind magnetizations $\sigma$. The forward shock is at $x\equiv \chi g =1$, the contact discontinuity is at $x=2$ (left vertical lines), the position of the  reverse shock depends on flow magnetization (right vertical lines). In all case the normalization of density $n$ is the same (it is an independent function that does not affect evolution of other quantities). In all cases kinetic pressure $f$ is zero on the inside, at $x=2_+$, while magnetic pressure $h^2/2$ balances the FS pressure $f(x=2_-)$. Generally, \Bf\ pressure increases towards the CD.}
 \label{xRSofsigma1}
\end{figure}

In Fig. \ref{xRSofsigma1} we compare the expected location of the shock for a given $\sigma$ with the one calculated the procedure described above.
 As expected, in the limit $\sigma \rightarrow 0$, $x_{RS} \rightarrow 4$. This good agreement serves as one of the test of our numerical procedures.
 
 Next, we need to determine  the wind density $n_w$ and \Lf\ $\gamma_w$. So far we have determined the location of the RS, $x_{RT}$, and the parameter $f_0$ (normalization of the kinetic pressure, as a function of wind magnetization. 
 The density at the RS  should equal the post-shock density,
 \be
 n_{RS}/n_{ext}= 2 n(\chi_{RS}) n_w \frac{\gamma_w} {\gamma_{RS}' }
 \ee
 (we assumed $u_w \sim \gamma_w   $ and similarly for  $u_{RS}$).
 Combining this with the expression for pressure (\ref{sigmaRS}), we find
 \ba &&
 \frac{n_w}{n_{ext}} = \frac{3 (\beta_{RS}' +1)^2 g_{{RS}} \eta _{\sigma } n_{{RS}}^2}{2 \beta_{RS}'   f_{{RS}} x_{{RS}}}
 \label{nnn} 
\\ &&
 \frac{\gamma_w}{\Gamma_{FS}}= \frac{4 \beta _{RS}'  \gamma_{RS}'   f_{RS} \sqrt{x_{{RS}}}}{3 (\beta _{RS}'+1) \sqrt{g_{{RS}}} \eta _{\sigma }
   n_{{RS}}}
   \label{ggg}
   \ea
To clarify, $\beta_{RS}',\,  \gamma_{RS}'  $ and  $\eta _{\sigma }$ are parameters determining the structure of the magnetized shock in it rest-frame, 
(K\&C, Eqns. (\ref{gammaRS}) and (\ref{sigmaRS})), parameters $x_{{RS}}$, $ g_{{RS}}$ and $f_{{RS}} $ (as well as $\sigma_w$) have been calculated previously.

   The relation for $n_{{RS}}$ involves the  integration  constant  $n_0$,  Eq. (\ref{111}).
  Then Eq. (\ref{nnn})  determines this integration constant in 
terms of the physical ratio ${n_w}/{n_{ext}}$. Since the constant $f_0$ is already determined by the magnetization, the 
 condition  (\ref{ggg}) then determines  ${\gamma_w}/{\Gamma_{FS}}$.

For example, in case of zero wind magnetization, $\sigma_w=0$, 
$
 \beta _{RS}' =1/3
,\,
 \eta_\sigma =1
 ,\,
 x_{RS}=4
 $, we find
 \ba &&
 \frac{n_w}{n_{ext}} = \frac{2 g_{RS} n_{RS}^2}{f_{RS}}
 \nn &&
  \frac{\gamma_w}{\Gamma_{FS}}= \frac{f_{RS}} {\sqrt{2 g_{RS} }n_{RS} }
 \ea
 where values  $n_{RS},\, f_{RS} $ and $g_{RS}$ are determined by (\ref{solutions}).

\begin{figure}[h!]
 \centering
 \includegraphics[width=.99\columnwidth]{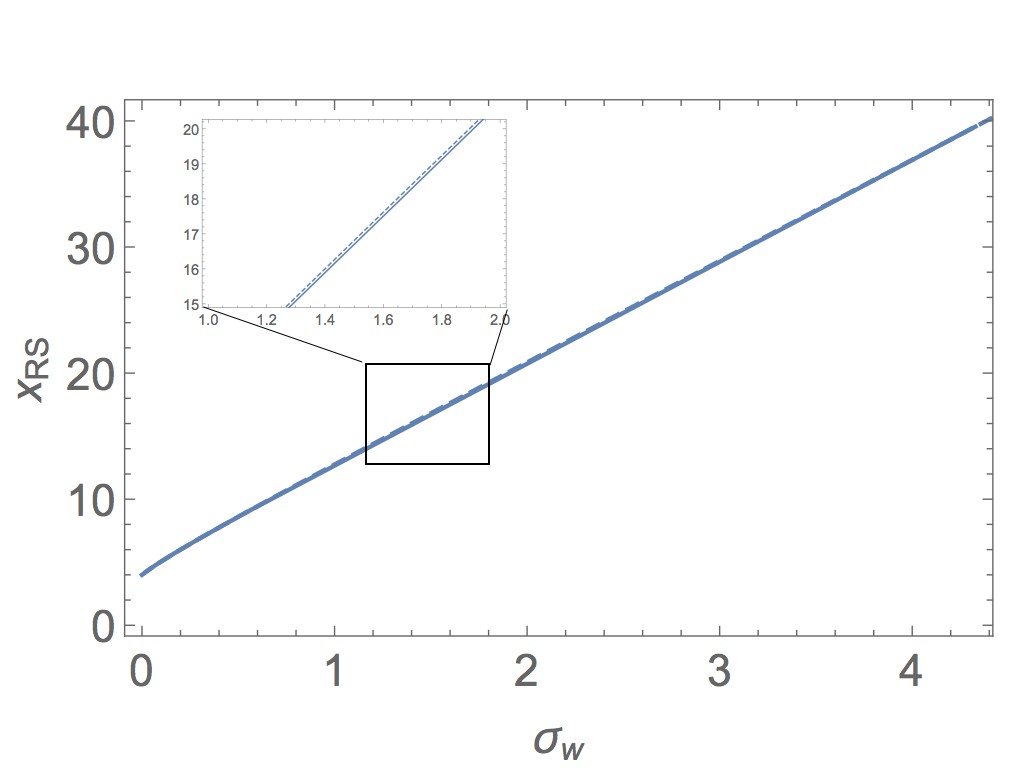}
 \caption{Comparison of the predicted locations of the RS in the self-similar coordinate $x$, Eq. (\protect\ref{xxRS}), dashed line, with the numerically calculated (solid line).  
}
 \label{xRSofsigma}
\end{figure}

In Fig.  \ref{xRSofsigma1} we plot flow variable $h,g,f$ and $h^2/2$.  Qualitatively,  for small wind magnetization, $\sigma_w \leq 1$, the \Bf\ within the shock wind flow increases towards the CD. For $m=1$ and $\sigma \geq 1$, the  \Bf\ and the \Lf\ remain nearly constant, as predicted by analytics,  Eq.
 (\ref{m1})

We also point out that for the simples case of $k=0, \, m=1$  Eqns. (\ref{16}) have a first integral
\ba && 
Z_3 = \log \left(\frac{1771561\ 11^{2/7} e^{\frac{x^2+2 x-8}{3 x}} x (x+2)^5}{16\ 2^{5/7} (7 x+8)^{44/7}}\right)+2
\nn &&
\frac{2 (x+2) \left(x^2+13 x-8\right) }{3 x (7 x+8)} f({Z_3})+{h}({Z_3})^2
-2 f(2_-)=0
\label{first}
\ea
We have verified that our numerical procedures conserve the  first integral (\ref{first}) to a precision of the order of  $\sim 1\%$.

\begin{figure}[h!]
 \centering
 \includegraphics[width=.99\columnwidth]{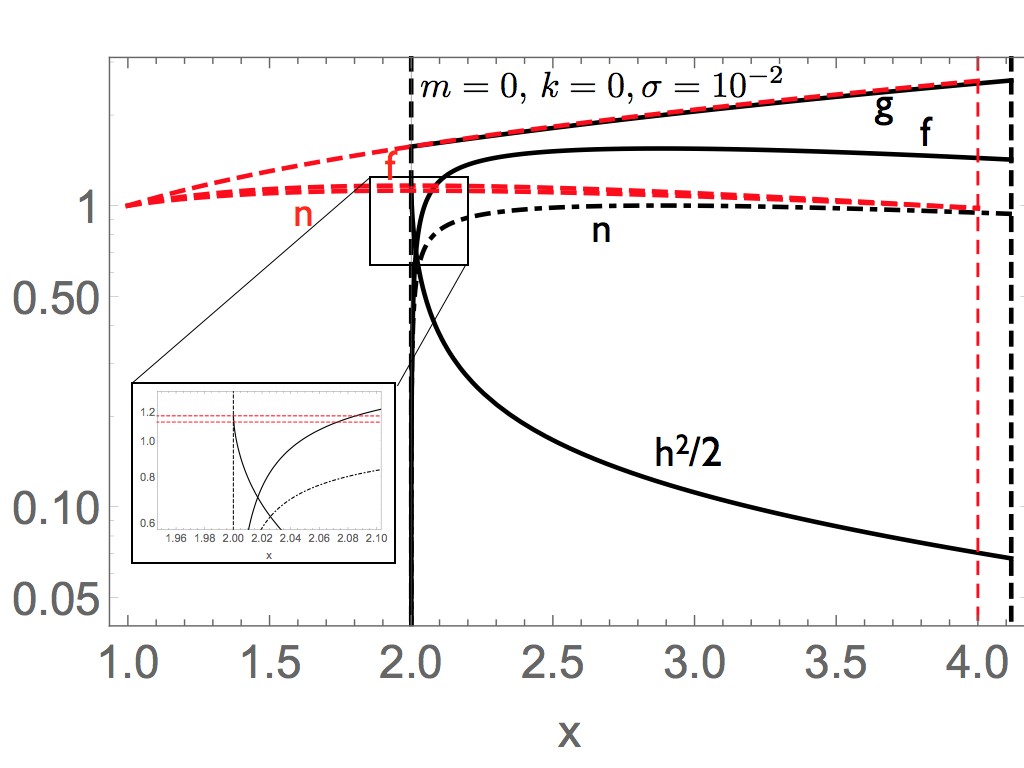}
 \caption{Example of continuous analytical solution for unmagnetized flow $m=0$, $k=0$ (red dashed curves) and the weakly magnetized shocked wind, $\sigma_w=10^{-2}$,
 $x_{RS}=4.11$.
 }
 \label{compa}
\end{figure}

\subsection{Magnetically-dominated RS}
As a special case, consider a highly magnetized ejecta. This could be  astrophysical interesting case, as  the energy supply in long-lived GBR central sources  is expected to be in a form of highly magnetized wind produced either by the central \BH\ or magnetar, \eg\  
\cite{2011MNRAS.413.2031M,2013ApJ...768...63L}. Neglecting kinetic  pressure, $f \rightarrow 0$, we find
 \ba &&
(\ln  n)' = -\frac{k (3 x-2)+2 m (x-2)-6 x+4}{8 (x-2) x}
 \nn &&
(\ln g)'=-\frac{k+2 m-2}{4 x}
 \nn &&
(\ln h)'=-\frac{k+2 m-2}{8 x}
 \ea
 with solutions
 \ba &&
 n \propto (2-x)^{\frac{2-k}{4}} x^{\frac{1}{8} (-k-2 m+2)}
 \nn &&
 g \propto x^{\frac{1}{4} (-k-2 m+2)}
 \nn &&
 h \propto x^{\frac{1}{8} (-k-2 m+2)}
 \label{m1}
 \ea
(In this case the FS is formally at infinity.) 
 For $m=1, k=0$ (the case of constant luminosity source in constant density environment) the  flow has constant \Lf\ and constant \Bf.   Our numerical results, Fig. \ref{xRSofsigma1}, are in agreement with this  limiting case.
         
         \section{Magnetized outside medium: formation of magnetosheath}
         \label{outs}
         The case of magnetized external medium was considered by \cite{2002PhFl...14..963L}. Let us briefly re-derive the main result here.
 In case of magnetized outer region, the CD is located at \cite{2002PhFl...14..963L}
 \be
 x\approx 2+ 6 \sigma_{ext}
 \ee
 with 
 \be
 h(1)= 2 \sqrt{2 \sigma_{ext}}
 \ee
 For consistency, instead of changing the $x$ coordinate of the CD, we shift a bit the location of the FS for the weakly magnetized external medium to
 $x_{FS}= 1-         6 \sigma_{ext}$.  
      
      Similar to the reverse shock flow (see also     \cite{2002PhFl...14..963L}),
         the pressure balance and continuity  require on the CD
\ba &&
f_- + h^2 _-/2=C_{CD}= f_+ + h^2 _+/2
\nn &&
g_-= g_+
\label{f41}
\ea
where $C_{CD}$ is some constant.

The system (\ref{18})  has an  integral of motion (if $k+m \neq 4$)
\be
f_\pm+{3 h_\pm^2/4}= (3/2) C_{CD}
\label{fff}
\ee

If we assume that on the each boundary a fraction $\alpha_\pm$ of pressure is contributed by kinetic pressure, 
\ba &&
f_\pm= \alpha_ \pm C_{CD}
\nn &&
h_ \pm = \sqrt{2 C_{CD} (1-\alpha_ \pm)},
\ea
then condition (\ref{fff}) and the pressure balance (\ref{f41}) require $\alpha_ \pm=0$   - kinetic pressure must vanish on both sides of the CD.

\begin{figure}[h!]
 \centering
 \includegraphics[width=.99\columnwidth]{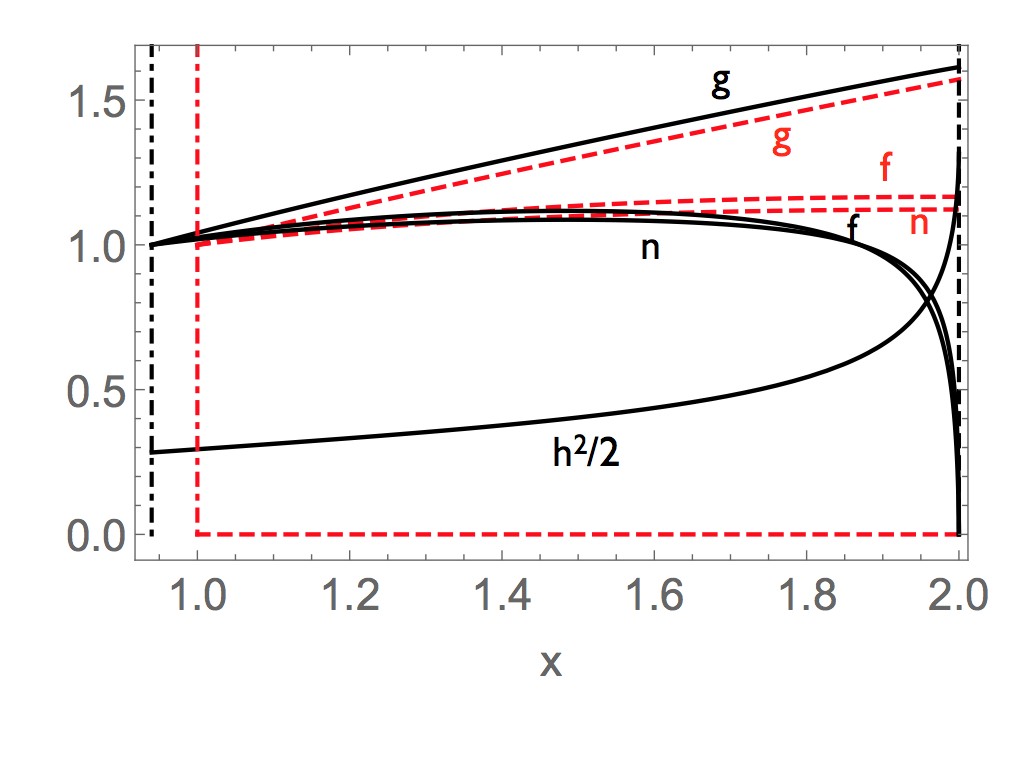}
 \caption{Structure of forward shock for $m=0$, $k=0$;  unmagnetized flow (red dashed curves) and the weakly magnetized external medium wind, $\sigma_{ext}=10^{-2}$.
The location of the CD is fixed at $x=2$; in the fluid case the FS is at $x_{FS}=1$, in the magnetized case it is at $x_{FS}=1-6 \sigma_{ext}$; $h(x_{FS}) =  2 \sqrt{2 \sigma_{ext}}$
 }
 \label{Plot-m-k-sigma017}
\end{figure}

Note a very different behavior of the density  on the CD in the fluid case, $n \propto (2-x)^{\frac{k-m}{m-12}}$ and with even small \Bf,   Eq. (\ref{111}). 
For example, in case of a wind and constant luminosity source ($k=2, \, m=0$) density becomes infinite on the  CD in the fluid case, $n \propto (2-x)^{-1/6}$, but is finite, $n \propto (2-x)^{0}$ in the magnetized case. Similarly, for $k=m=0$ density is finite on the CD in the fluid case, but becomes zero for the (weakly) magnetized case, Fig. \ref{Plot-m-k-sigma017}.

\begin{figure}[h!]
\includegraphics[width=\linewidth]{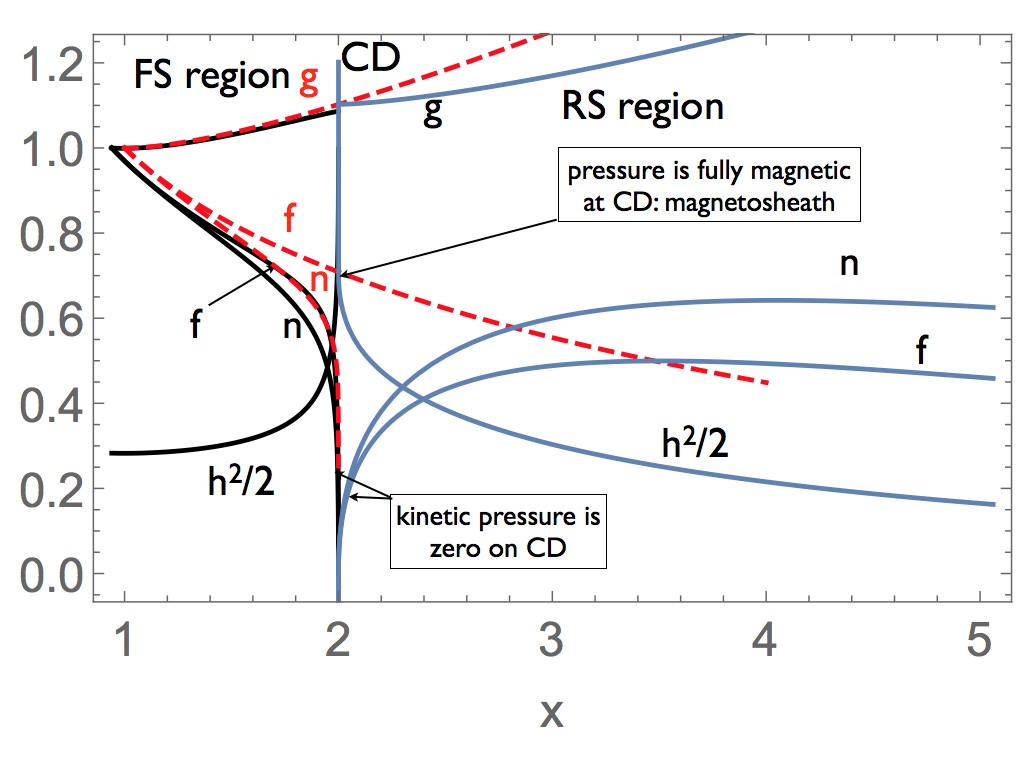}
\caption{Self-similar structure of magnetized flow between  the FS and the RS for constant power source in constant density environment ($m=1$, $k=0$) in coordinate $x= \chi g$. Location of the CD is fixed at $x=2$. Red dashed lines are solution for fluid case, $h=0$, solid black is for magnetized case in the FS region, solid blue is for magnetized  RS flow.  In this case  the FS is located at $x=1$, RS at $x=4$, density is zero on the CD, while the \Lf\ and the pressure are continuous. Solid lines are the solution for magnetized FS and RS flow. FS is located at $x< 1$.  External magnetization is $\sigma =10^{-2}$, the wind magnetization is $\sigma_w =0.1$.  
Though magnetization parameter is small in the bulk of both the FS and the RS flows, it is dominant on the CD - kinetic pressure is zero on both sides of the CD. In the RS region the normalization of density is arbitrary. Small mismatch of the \Lf\ on the CD is a numerical artifact  of locating the FS.
}
\label{Plotm1k0BothSides}
\end{figure}

The full structure, magnetized FS and RS, for the case of constant wind in constant external medium, $m=1$, $k=0$, is pictured in Fig. \ref{Plotm1k0BothSides}. Importantly, the \Bf\ dominates over the plasma pressure on both sides of the CD. 

Finally, let us comment on the relative behavior of the density and the \Bf\ on the CD. Neglecting effects of the magnetic pressure, $h^2 \rightarrow 0$, we find that close to the CD, at  $g \chi =2$, the evolution of density and \Bf\ obeys
\be
\partial_\chi \ln \frac{ n}{h} =  \frac{m-k}{k+m-4},
   \ee
   Thus, only for positive $m-k$  the density increases with increasing \Bf.

\section{Reconnection at the magnetosheath}
\label{magnetosheath}

\subsection{The role of reconnection in astrophysical high energy sources.}
 Recently, a number of observations  question the dominant paradigm of shock acceleration in astrophysical high energy sources, like Pulsar Wind Nebulae (PWNe), Active Galactic Nuclei (AGNe) and Gamma Ray Bursts (GRBs). Particularly important was a detection of Crab flares   \citep{2011Sci...331..739A,2011Sci...331..736T,2012ApJ...749...26B}. The Crab  Nebula is the paragons of other astrophysical sources -   models of  AGNe and GRBs use the unipolar inductor paradigm - they are 
 driven by highly magnetized rotating compact sources \citep[\eg][]{BlandfordZnajek,Usov92,2006NJPh....8..119L,2007MNRAS.382.1029K,2004ApJ...611..380T,2003ApJ...597..998L}. 
 The  reconnection models developed for  Crab flares  might/should be applicable to 
magnetized jets of  AGNe and GRBs.

Magnetic reconnection can lead to explosive release of magnetic energy, e.g. in solar flares.  However, properties of plasma in the Crab Nebula, as well as magnetospheres of pulsars and magnetars, pulsar winds, AGN and GRB jets and other targets of relativistic astrophysics,  are very different from those of more conventional Solar and laboratory plasmas \citep{2013SSRv..178..459L}.  The physics of particle acceleration in relativistic current sheets has been addressed in a number of recent studies \citep[\eg][and others]{2012ApJ...750..129B,2015ApJ...806..167G,2015ApJ...805..163D,2016arXiv160305731L}.
All these work explore somewhat different aspects of reconnection in highly magnetized plasma, but they generally agree on the following points: (i) in nearly ideal plasma the current sheet becomes unstable to formation of plasmoids  \citep{2007PhPl...14j0703L,2010PhRvL.105w5002U}
 that are ejected with near-\Alfven velocity - in highly magnetized plasma $\sigma \geq   1$ the ejection can be relativistically fast;
(ii) reconnection does produce power law distribution of accelerated particles, with the spectrum reaching $p=1$ for highly magnetized set-up  $\sigma\gg    1$ \cite{guo_14,werner_16,2016arXiv160305731L}.

\subsection{Magnetosheath - formation of magnetized depletion layer near the CD}

Creation of highly magnetize layer near a CD (the magnetopause) discussed in this work is not specific to GRB outflows - it is a well known phenomenon in space physics \cite{1966P&SS...14..223S,1967P&SS...15..239A}. In the highly magnetized region plasma density is depleted \cite{1976JGR....81.1636Z}. There are evidence of magnetic reconnection   in the magnetosheath between the Solar wind and the Earth's \Bf. In astrophysical applications   \cite{1965ApJ...142..491K} considered a supernova explosion in magnetized mediums and concluded that  near the CD the ``magnetic field is large no matter how small its interstellar value" \citep[see also][]{2004ApJ...609..785L,Lyutikovdraping}

 In case of GRBs the formation of highly magnetized boundary layer in the FS region has been discussed  previously by   \cite{2002PhFl...14..963L}; in this paper  we demonstrated that in the RS region the pressure is mediated by the \Bf\ as well. Thus, the kinetic pressure becomes zero on both sides of  the CD no matter how small the \Bf\ is in the bulk. It is the thickness of the magnetosheath that depends on the  \Bf\ in the bulk.

\begin{figure}[h!]
 \centering
 \includegraphics[width=.99\columnwidth]{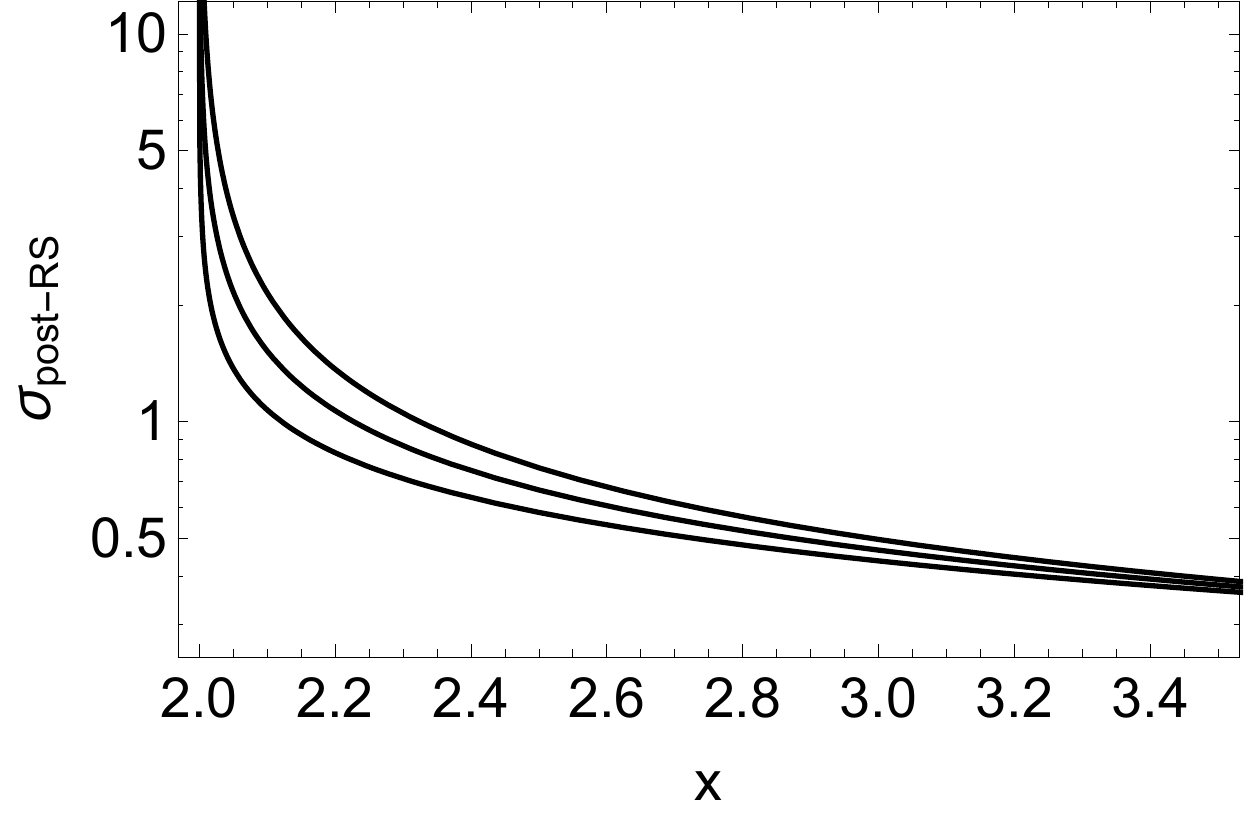}
 \caption{Local magnetization in the shocked wind flow, starting  with $\sigma_w=0.1$ for $k=0$,  $m=0,1,2$ (top to bottom)}
 \label{sigmaofx}
\end{figure}

 Formally, the magnetization becomes infinite on the CD, Fig. \ref{sigmaofx}. We expect that plasma instabilities will limit. Since plasma in relativistic flows is expected to be highly collisionless, relativistic double-adiabatic theory is more relevant near the CD
 \cite{1991PhFlB...3.1871G}. It is expected that  the development of  firehose and mirror   instabilities \cite{1993PhRvE..47.4354G}  will heat the plasma and will limit $\sigma \leq  $ few.

\subsection{Reconnection at the magnetosheath: energetic considerations}

 The \Bfs\ in the external medium and in the wind are generally not aligned. Thus the CD become a tangential discontinuity, where the \Bf\ experiences a jump in direction. The plasma is electron ion on the FS side, while it can be purely pair plasma on the RS side.
Since   near the CD the \Bf\ is amplified while plasma is heated to the state of near equipartition $ w  \sim B^2/(8\pi)$ (where $w$ is rest-frame enthalpy), we expect that the 
 tangential discontinuity has a size of the order of the Larmor radius  and of the skin depth.  We expect that highly collisional  plasmas become highly dissipative when the field changes on the scale of Larmor radius  and/or  skin depth \cite{guo_14}. This will set-up a reconnection layer on the CD. The reconnecting plasma has $\sigma \sim $ few on both side, but one side is electron-ion while the other could be a pair plasma. As reconnection proceeds, more magnetic energy is brought to the reconnection layer. The  reconnection layer is expected to break into plasmoids \cite{2007PhPl...14j0703L,2010PhRvL.105w5002U}, which will be ejected ejected with mildly relativistic velocities.

Let's estimate how much magnetic energy is available near the CD. Since we consider  reconnection  between the external and internal media, the amount of energy available will be determined by the smaller flux (between the wind and the swept-up material). 
The magnetic flux  produced by a central source with luminosity $L_w$ in time t, $ \Phi_{B_s} \sim \sqrt{L_w c} t$, is typically much larger than the swept up magnetic flux.
Thus, energy budget is determined by the external flux. Let us next estimate the magnetic energy in the swept-up material.
If the external \Bf\ is $B_{ex}$, then at time $t$ the total swept-up
 flux is  
 \be
 \Phi_{B_{ex}} = \pi (c t)^2 B_{ex}
 \label{swept}
 \ee

Near the CD \Bf\ is highly amplified. It is expected that the development of instabilities in the collisionless plasma (mirror, firehose, and cyclotron) will limit the local plasma magnetization to $\sigma \sim $ few. Let us assume that the  compression of the \Bf\ saturates at equipartition. 
The equipartition field in the flow frame is 
\be
B_{eq} '\sim \Gamma \sqrt{32 \pi c^2  {\rho_ {ext}}};
\ee
It is $\Gamma$ times higher in the lab frame
\be
B_{eq} \sim\Gamma ^2  \sqrt{32 \pi c^2  {\rho_ {ext}}}=  3 \times 10^{4} {\rm G} \sqrt{n_{ex}} \left( \frac{\Gamma }{300}\right)^2
\ee
Equating the swept-up flux (\ref{swept}) with the flux through a ring of radius $( ct )$  and thickness $\Delta_r$, we estimate the thickness of the magnetosheath in lab frame
\ba &&
2 \pi (c t) \Delta_r B_{eq}=  \Phi_B 
\nn &&
  \frac{\Delta_r }{ct} = \frac{\sqrt{\sigma_{ex}}}{4 \sqrt{2} \Gamma^2}
  \nn && 
    \frac{\Delta_r }{ct_{ob}} = \sqrt{\frac{\sigma_{ex}}{16 }}
\ea
where  $\sigma_{ex}=B_{ex}^2/(4 \pi \rho_{ex} c^2)=5 \times 10^{-11} B_{-6}^2  n^{-1}$.
Thus, after $\sim 100 $ seconds of observer time all the swept-up magnetic flux is connected in layer of  only $\approx 7.5 \times 10^6$ cm. 
(In the flow frame it is $ \Gamma$ larger.)

Total magnetic energy per  solid angle $d  \Omega/(4\pi)$  at time $t_{ob}$ in our frame is
 \be
E_B= 4 \pi (  ct)^2 \Delta_r  \frac{B_{eq} ^2}{8\pi} \frac{d \Omega}{4\pi}= 
10^{53} {\rm erg} \left( \frac{B_{ex} }{10^{-6} {\rm G}}\right) 
\left( \frac{\Gamma }{300}\right)^8 \left( \frac{t _{ob}}{10^{4} {\rm s}}\right)^3
\left( \frac{d \Omega }{10^{-3} }\right) 
\propto t_{ob}^{1/3}
\label{EB1}
\ee
where the last approximation assumes that the blast wave is powered by a constant luminosity source.
Note that energy (\ref{EB1}) is larger than what a simple boost would have produced
 \be
E_{B,0}\sim \frac{4 \pi}{3} \Gamma^2  \frac{B_{ex} ^2}{8 \pi} ( ct)^3 \frac{d \Omega}{4\pi} = 2 \times 10^{47} {\rm erg},
\ee
due to the fact that \Bf\ is dynamically compressed at the CD.

Thus, there is enough magnetic flux swept-up to account for energetics of $X$-ray afterglows and GeV emission.

\subsection{Reconnection events in the magnetoheath}
 
\subsubsection{Quick X-ray flares from mini-jets}
One of the most surprising results of {\it Swift}  observations of early afterglows, at times $\leq$ 1 day, is the ubiquitous presence of  flares 
\cite[\eg][]{2006ApJ...642..389N,2013FrPhy...8..661G,2016ApJ...829....7L}.  The flares are very short, with the duration much shorted that the observer time since the explosion $t_{ob}$.,
$\Delta t \sim 0.1 t_{ob}$. This presents  a challenge to the models, since even if  emission generation from a  relativistic spherical shell   located at radius $R=  c t_{em}$ and propagating with \Lf\ $\Gamma$ 
 is switched-off instantaneously, the time delay for photons emitted at angles $\theta \sim  1/\Gamma$ will be  
 \be
 \Delta t \sim (R/c) / \Gamma^2 \approx t_{ob}. 
\label{Deltat}
\ee

\begin{figure}[h!]
 \centering
 \includegraphics[width=.99\columnwidth]{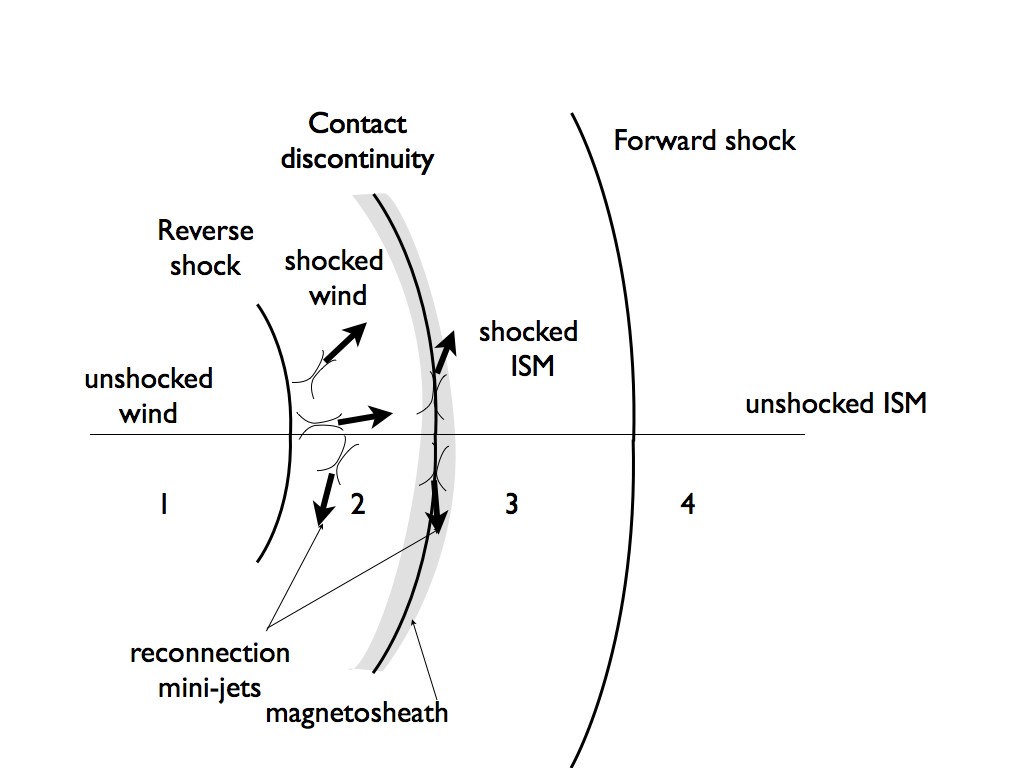}
 \caption{A cartoon of Crab-like flares/mini-jets produced by reconnection events in the highly magnetized post-RS flow. Near the CD a highly magnetized magnetosheath forms on both side.}
 \label{mini}
\end{figure}

Relation (\ref{Deltat}) also offer a possible resolution of the problem of short flare duration \citep[see][]{Lyutikov:2006a}, see Fig. \ref{mini}. If an emitting region is moving with bulk \Lf\ $\Gamma$, the observer time at emission time $t_{em}$ is 
$t_{ob} \sim t_{em}/\Gamma$. But, if the emission region is moving with respect to the bulk frame with $\gamma_j \geq 1$, the effective \Lf\ in the observer frame is 
\be
\Gamma_{em} \approx 2 \gamma_j \Gamma
\ee
Thus, the observed duration is
\be
 \Delta t \sim (R/c) / \Gamma_{em}^2\approx  \frac{t_{ob}}{4 \gamma_j ^2} \ll t_{ob}
 \ee
 This was one of the key points of the mini-jet model of \cite{Lyutikov:2006a}, that observed duration can be much shorter even for mild internal {\Lf}s \citep[see also][]{2009MNRAS.395L..29G,2011MNRAS.410.2422B}.

 \cite{2011MNRAS.415.2081C}  calculated  statistical properties of emission from mini-jets.
 It was found  that the distribution of observed fluxes $f(F)$ follows a power law $f(F) \propto F^{-(q+1)/q} dF \approx  \propto F^{-1} dF$, where $q=3+\alpha$ ($F_\epsilon \propto \epsilon^{-\alpha}$ is the Doppler boosting factor from the emission rest-frame to the observer frame \citep{Lind:1985}; this is a known result in the theory of AGN jets \citep{1984ApJ...280..569U}). Thus, rare bright events contribute significantly to overall emission.
 
Reconnection events in highly magnetized plasma can be very efficient in accelerating non-thermal particles \citep[\eg][]{2016arXiv160305731L}. Acceleration events can proceed explosively, with exceptionally fast acceleration rates (so that the accelerating \Ef\  is of the order of the \Bf), with particles reaching the radiative reaction limit.
 
\subsubsection{GRBs and Fermi LAT photons: a chance for synchrotron}

The detection of GRBs by the \Fermi\ satellite \citep{2009Sci...323.1688A} is an important probe of  GRB physics. The recent observation of the 95 GeV photon (125 GeV in rest frame) at $\sim 250$ seconds and 30 GeV  (40 GeV in rest frame)  at $\sim 35$ ksec  from GRB130427A  \citep{2014Sci...343...42A,2014Sci...343...51P} is problematic if one tries  to related the  origin of LAT photons to  synchrotron 
emission \citep{2013arXiv1306.5978L}.
There is a {\it acceleration theory-independent}  upper limit on the frequency of synchrotron emission by radiation reaction-limited  acceleration of electrons \citep{2010MNRAS.405.1809L}. In astrophysics the effective  accelerating \Ef\ is a fraction $\eta \leq 1$ of the \Bf\ (this is equivalent to  acceleration on time scale of inverse cyclotron frequency $1/( \eta \om_{B,rel}) $, where $\om_{B,rel}=\gamma/\om_B$ is relativistic cyclotron frequency of a particle). Equating the acceleration rate and synchrotron energy losses,
\be
\eta e B c \sim  {e^2 \over c}  \gamma^2 \om_B^2
\label{eta}
\end{equation}
the peak energy of  synchrotron emission is then
\be
\epsilon_{\rm max} ' \sim  \hbar  { m c^3 \over e^2} \approx 100 \mbox{ MeV}.
\label{emax}
\end{equation}
Note, the upper limit (\ref{emax}) assumes the most efficient, {\it non-stochastic}, DC-type acceleration. 

If the emitting plasma moves with a Lorentz factor $\Gamma$ towards the observer, the observed maximal frequency is $\sim 2 \Gamma \epsilon_{\rm max}$. 
The Fermi photons come over times much longer than the duration of the prompt emission. Assuming the \cite{BlandfordMcKee} scaling of the Lorentz factor, we find
\be
\Gamma \sim\left(  {E_{iso} (1+z)\over c^5 t_{ob}^3 m_p n_{ISM} } \right)^{1/8}, \, \epsilon_{\rm max} = 2 \Gamma \epsilon_{\rm max} ' / (1+z)
\label{Emax}
\end{equation}
where we introduced the cosmological factor $(1+z)$.
The relation (\ref{Emax}) puts  a constraint on the maximal synchrotron energy emitted at the observer time $t_{ob} $. The burst GRB130427A has particular tight constraints. If emission  comes from the FS, then assuming  $E_{iso} =10^{54}$ erg and $ n_{ISM} =1$, the maximum photon energy at $t_{ob} =3.5 \times10^4 $ sec is then $\sim 12$ GeV. 
This is nearly 3 times smaller than the observed photon energy. Also notice that $ \epsilon_{\rm max} $ is very insensitive to the precise values of $E_{iso} $ and $ n_{ISM} =1$, $\epsilon_{\rm max}  \propto (E_{iso}/ n_{ISM} )^{1/8}$. Thus, in the FS model either the isotropic equivalent t energy should be $3^8 \approx  6 \times 10^3$ times higher or density correspondingly lower. These are unreasonable parameters (recall  that this also assumes absolute maximum for the efficiency of acceleration, $\eta = 1$ in Eq. (\ref{eta})).

\begin{figure}[h!]
\includegraphics[width=\linewidth]{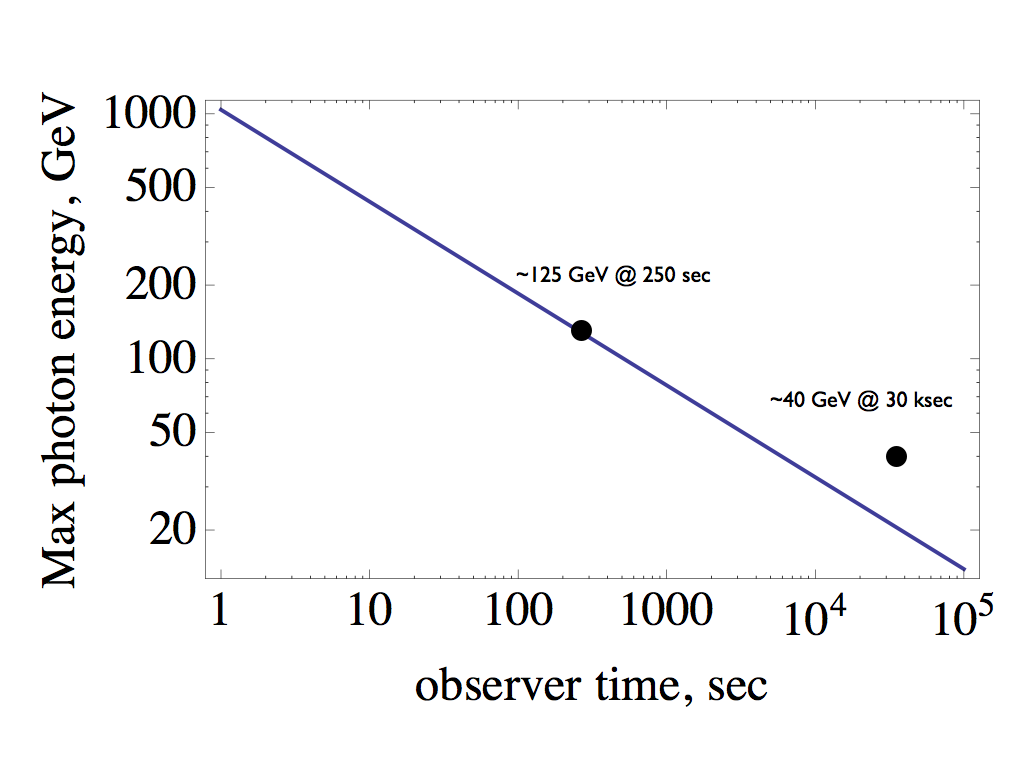}
\caption{Upper limit of synchrotron emission from the forward shock and two LAT photons from  GRB130427A. The very late high energy photon, $\sim 35 GeV$ at $\sim 35 $ ksec violates the upper limit on the synchrotron emission.}
\label{GRB-Upper-Sinch}
\end{figure}

In contrast to the above limitations on synchrotron spectrum, \cite{2013ApJ...779L...1K}  established that late (after few days) $X$-ray spectrum form a single continuous afterglow spectral component from optical to multi-GeV.  \cite{2013ApJ...779L...1K}  argued  the emission mechanism is synchrotron, which is at odds with the above considerations.

If we take arguments of  \cite{2013ApJ...779L...1K} that continuation of the spectrum implies synchrotron mechanism, the photon energy excludes the FS. The mini-jet model can come to rescue:   in the mini-jet model in order  to produce a 30 GeV photon it is
required that $\gamma_j \sim  2$. Such {\Lf}s  are not expected in the weakly magnetized  FS region, where the sonic \Lf\ is only $\sqrt{3/\sqrt{8}}= 1.06$. As we discussed, reconnection in the mildly magnetized post-RS flow and/or at the magnetosheath, \S \ref{magnetosheath} can produce such motions.

         \section{Conclusion}
         In this paper
   we considered the self-similar        
   double-shock structure of relativistic magnetized outflows. Dynamically connecting outside region (the forward shock flow) with the inside 
   region (the reverse shock flow) requires resolving flow singularities at the contact discontinuity. We find that for most astrophysically interesting cases the \Bf\ in the reverse shock region fully compensates the force balance across the contact discontinuity: density and kinetic pressures are zero on the inside of the CD (and become so in a non-analytic way). Thus, density and  pressures  in the reverse shock region are not related to the forward shock region. In order to find the asymptotic scalings of pressure and density in the vicinity of  the contact discontinuity we devised a scheme that allows us to calculate the flow properties behinds the reverse shock as functions of wind power, density and magnetization. 
   
 We then discussed a possibility  that particles emitting early $X$-ray afterglows,  as well as Fermi GeV photons, are accelerated via magnetic reconnection processes in the post-reverse shock  region of long-lasting central engine. $X$-ray plateaus are produced by the quasi-steady reconnection (or by a collection of  mini-flares), while the 
 $X$-ray flares are produced by explosive reconnection events, akin to Crab Nebula flares.
 
  Previously, \cite{2017ApJ...835..206L} discussed a possibility that early $X$-ray afterglows are produced in the reverse shock region of the long-lasting wind. 
   The model  of  \cite{2017ApJ...835..206L}  reproduces, in a fairly natural way,  the overall trends  and yet  allows for variations in the temporal and spectral evolution of early optical and $X$-ray afterglows. The mechanism of particle acceleration was assumed to be at the reverse shock - here we propose acceleration via reconnection events in the shocked wind region. Since most of the observed properties of the emission are dominated by relativistic kinematic effects, results of  \cite{2017ApJ...835..206L}  are generally applicable to the present model as well.
 
  Especially important for reconnection events  could be the region near the contact discontinuity, where
 a highly magnetized region, magnetosheath,  is formed on both sides. Very high energy GeV photons could be of synchrotron origin, produced in relativistically moving exhaust mini-jets with internal  \Lf\  $\sim $  a few.

I would like to thank Maxim Barkov, Rodolfo Barniol Duran, Dimitrios Giannios and Juan Camilo Jaramillo   for discussions. 

This work had been supported by   NSF  grant AST-1306672 and DoE grant DE-SC0016369.

        \bibliographystyle{apsrev}
\bibliography{/Users/maxim/Home/Research/BibTex}

 \appendix
 
\section{Unmagnetized case - exact solutions}

Self-similar solution for the unmagnetized case are well known (B\&M), but 
 the possibility of extending these solutions to the RS region has not been discussed - this is a somewhat tricky issue as we discuss next.
 In the fluid  case  equations (\ref{16}) simplify (B\&M)
\ba
&&
{ \partial \ln f \over  \partial x}= 
\frac{m
   (x-8)-4 x+8}{m (x-4)+x^2+12 x-4}
   \nn &&
{ \partial \ln g \over  \partial x}= 
\frac{m (x-7)+2 (x+2)}{m (x-4)+x^2+12 x-4}
\nn &&
 { \partial \ln n \over  \partial x}=
\frac{m \left(x^2-11 x+22\right)-6 (x-2)^2}{2 (x-2) \left(m (x-4)+x^2+12 x-4\right)}
\label{17}
\ea
Close to   the CD, $x \rightarrow 2_ -$,  we find
\ba
&&
 { \partial \ln f \over  \partial x}= 
 -   \frac{3m}{12-m}
  \nn && 
 { \partial \ln g \over  \partial x}= 
    \frac{8-5m}{2(12-m)}
 \nn &&
 { \partial \ln n \over  \partial x}=
    \frac{m}{(12-m)(x-2)}
\label{181}
\ea

Equations (\ref{181}) clearly demonstrate that the CD is the special point of the flow. 
 There is no singularity in the \Lf\  $g$ or pressure $f$, so that the two flow can always connect smoothly velocity-wise. 
The equation for the density can have singularity at the CD for  $m\neq 0$.

Equations  (\ref{17})  can be integrated for $m\neq 3$ ($m=3$ is the impulsive solution with no CD):
\ba &&
n=(2-x)^{-\frac{m}{m-12}}  { Z_1}^{\frac{m}{4}+\frac{6}{m-12}-1}  { Z_2}^{-\frac{m^3+10 m^2-416
   m+1152}{4 (m-12) \sqrt{m^2+40 m+160}}}
   \nn &&
   f= { Z_1}^{\frac{m-4}{2}}  { Z_2}^{-\frac{m (m+24)-64}{2 \sqrt{m (m+40)+160}}}
   \nn &&
   g= { Z_1}^{\frac{m+2}{2}}  { Z_2}^{-\frac{m (m+28)+16}{2 \sqrt{m (m+40)+160}}}
   \nn &&
   Z_1= \frac{(m+12) x-4 (m+1)+x^2}{3 (3-m)}
   \nn &&
   Z_2=\frac{\left(1+\frac{m+14}{\sqrt{m^2+40 m+160}}\right) \left(1-\frac{m+2 x+12}{\sqrt{m^2+40
   m+160}}\right)}{\left(1-\frac{m+14}{\sqrt{m^2+40 m+160}}\right) \left(1+\frac{m+2 x+12}{\sqrt{m^2+40
   m+160}}\right)}
   \label{solutions}
   \ea
   Solutions (\ref{solutions})  and the definition $\chi= x/g$ represent exact analytical solutions for the structure of ultra-relativistic fluid shock waves.
 Note that the CD, located at  $x=2$, is the special point of the density structure, but not pressure or \Lf. 
 Only in the special case of $m=0$ the density  can be continuos on the CD
 

Note that addition of a weak, dynamically unimportant \Bf\ breaks the self-similar solutions  (\ref{solutions}) even for $m=0$. For weak \Bf, neglecting terms $\propto h^2$,  the \Bf\ evolves according to 
\be
h \propto (2-x)^{\frac{4-m}{m-12}}  {Z_1}^{\frac{m^2-14 m+40}{4 (m-12)}} {Z_2}^{-\frac{m^3+12
   m^2-384 m+704}{4 (m-12) \sqrt{m^2+40 m+160}}} 
\ee
with the derivative 
$
   h' \propto (2 - x)^{-2 (8 - m)/(12 - m)},
$
diverging on the CD

\section{Special case $m=2$, $k=2$}
\label{m2k2}

\begin{figure}[h!]
 \centering
 \includegraphics[width=.49\columnwidth]{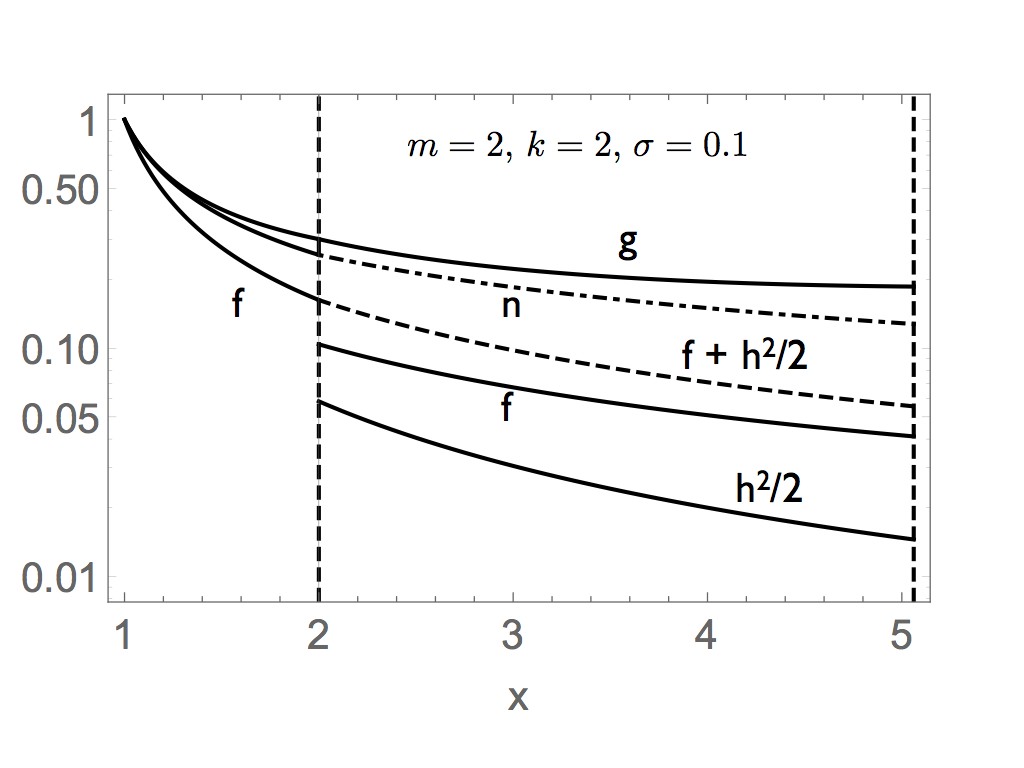}
  \includegraphics[width=.49\columnwidth]{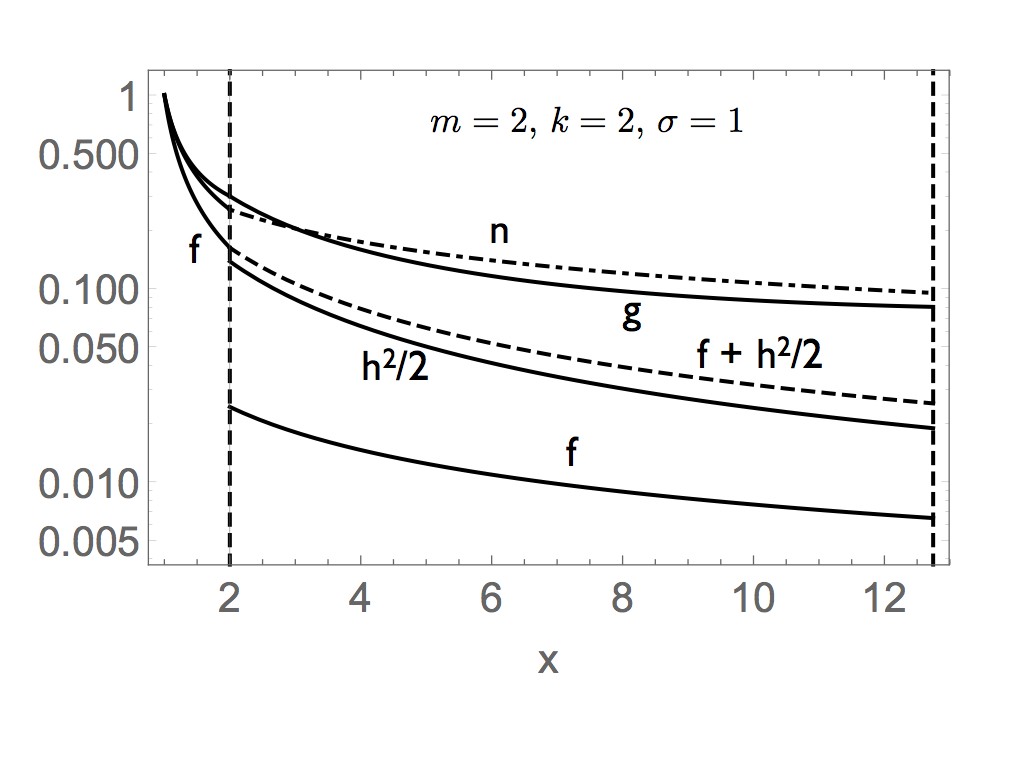}\\
   \includegraphics[width=.49\columnwidth]{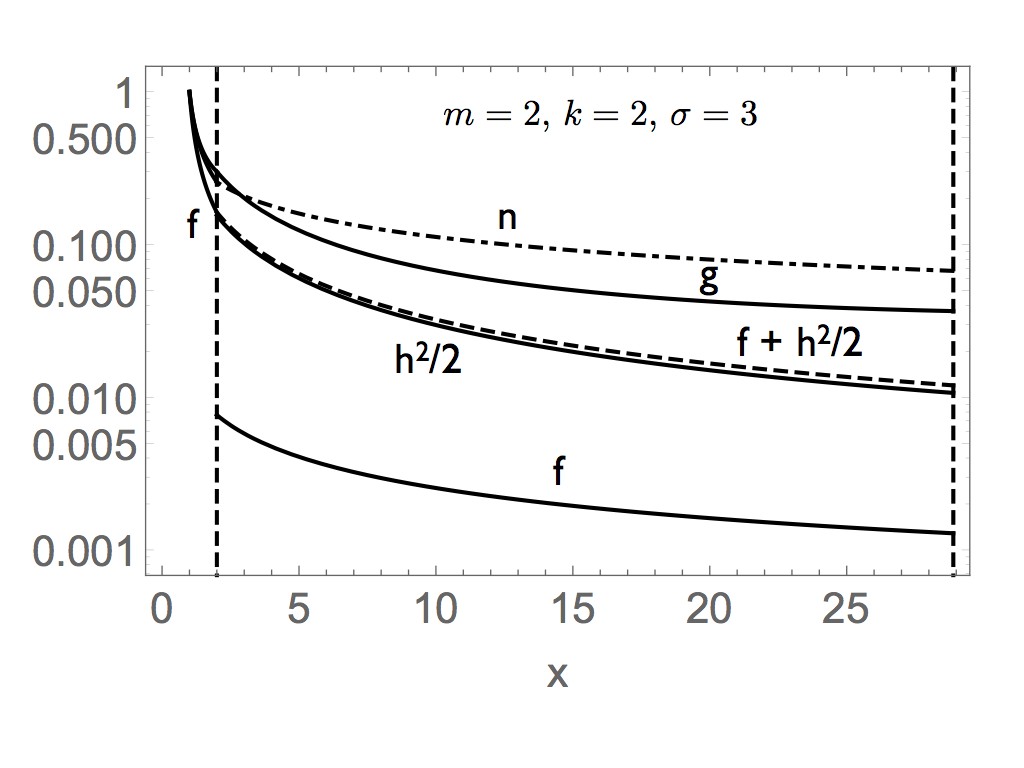}
    \includegraphics[width=.49\columnwidth]{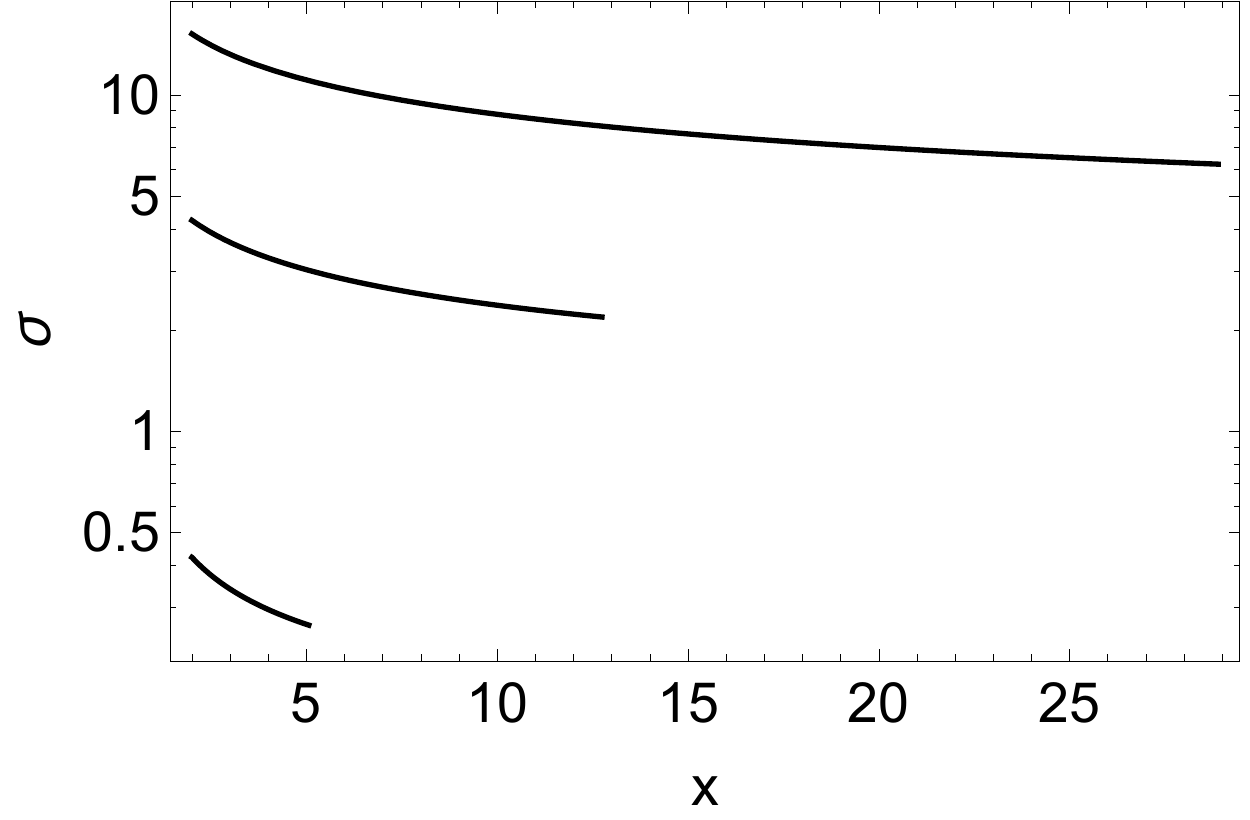}
 \caption{Flow functions and magnetization for the special case $m=2, \, k=2$. Normalization of density (dot-dashed line) is arbitrary at $x>2$. Dashed line is the total pressure, $ f+h^2/2$. }
 \label{Plotm2k2001}
\end{figure}

As we discussed above, the case $m=2$, $k=2$ is a special one - both  the kinetic pressure $f$ and \Bf\ can remain finite on the CD, with the ratio depending on external magnetization. 
In this case  equations (\ref{16}) take the form 
\ba &&
 (\ln  n)' =-\frac{3  \left(8 f+3 h^2\right)}{2 \left(f \left(x^2+14
   x-12\right)+9 h^2 x\right)}
   \nn &&
 (\ln  f)' =-\frac{2  \left(8 f+3 h^2\right)}{f
   \left(x^2+14 x-12\right)+9 h^2 x}
  \nn &&  
  (\ln g)' =\frac{ \left(4 f (x-4)-9
   h^2\right)}{f \left(x^2+14 x-12\right)+9 h^2 x}
     \nn &&
   (\ln h)' =-\frac{3  \left(8 f+3
   h^2\right)}{2 \left(f \left(x^2+14 x-12\right)+9 h^2 x\right)}
   \ea
  The  CD is not a a special point, hence solutions can be  extended throughout the CD. 
  Following the prescription in \S \ref{outs}, for a given ration of pressures on the CD, 
  $2 f(2_+)/h^2(2_+)=\sigma/(1-\sigma)$ we integrate the equations into the RS region, find the point where local magnetization matches the RS shock conditions - this determines 
  $\alpha (\sigma_w)$. In this case the density is not required to be continuos - it ca experience a jump at the CD. The functions $f$ and $n$ are simplify related to $h$:
  $\propto h^{4/3}$, $n \propto h$, see Fig. \ref{Plotm2k2001}. No that in all case magnetization increases towards the CD, Fig. \ref{Plotm2k2001}.
     
\end{document}